\documentclass[12pt]{iopart}
\usepackage{iopams,psfig,color,harvard,graphicx}

\newcommand{\z}[1]{{\tt z#1}}
\newcommand{\zABI}[1]{\textcolor{red}{ABI #1}}
\renewcommand{\z}[1]{}
\renewcommand{\zABI}[1]{}

\newcommand{\Secref}[1]{\Sref{#1}}
\newcommand{\secref}[1]{\sref{#1}}

\newcommand{\Eqref}[1]{\Eref{#1}}
\newcommand{\eqref}[1]{\eref{#1}}
\newcommand{\eqsand}[2]{\eref{#1} and \eref{#2}}
\newcommand{\Figref}[1]{\Fref{#1}}
\newcommand{\figref}[1]{\fref{#1}}

\newcommand{\Tabref}[1]{\Tref{#1}}
\newcommand{\tabref}[1]{\tref{#1}}

\newcommand{\bea}{\begin{eqnarray}}
\newcommand{\eea}{\end{eqnarray}}
\newcommand{\bal}{\begin{aligned}}
\newcommand{\eal}{\end{aligned}}
\newcommand{\bga}{\begin{gathered}}
\newcommand{\ega}{\end{gathered}}

\newcommand{\lt}{\left}
\newcommand{\rt}{\right}
\newcommand{\la}{\langle}
\newcommand{\ra}{\rangle}

\newcommand{\const}{{\rm const}}
\newcommand{\eps}{\varepsilon}

\newcommand{\dd}{\partial}
\newcommand{\vdel}{\bnabla}
\newcommand{\vu}{\bi{u}}
\newcommand{\vf}{\bi{f}}
\newcommand{\vB}{\bi{B}}

\newcommand{\vk}{\bi{k}}

\newcommand{\lf}{L}
\newcommand{\lvisc}{l_\nu}
\newcommand{\lres}{l_\eta}

\newcommand{\nueff}{\nu_{\rm eff}}

\newcommand{\du}{\delta u}
\newcommand{\dB}{\delta B}
\newcommand{\dvB}{\delta\vB}
\newcommand{\umf}{U}
\newcommand{\urms}{u_{\rm rms}}
\newcommand{\dBrms}{\delta B_{\rm rms}}
\newcommand{\Bo}{B_0}

\newcommand{\usq}{\la |\vu|^2\ra}
\newcommand{\Bsq}{\la |\vB|^2\ra}
\newcommand{\dBsq}{\la|\dvB|^2\ra}

\newcommand{\Pm}{{\it Pm}} 
 
\renewcommand{\Re}{{\it Re}} 
\newcommand{\Rel}{\Re_\lambda} 
\newcommand{\Rm}{{\it Rm}} 
\newcommand{\Rmc}{\Rm_{\rm c}} 
\newcommand{\Rmcmax}{\Rm_{\rm c}^{\rm(max)}} 
\newcommand{\Rmcinf}{\Rm_{\rm c}^{(\infty)}} 

\newcommand{\gammainf}{\gamma_\infty} 

\newcommand{\tcorr}{\tau_{\rm corr}}

\begin{document}


\title[Fluctuation dynamo and turbulent induction at low magnetic Prandtl numbers]{Fluctuation dynamo and turbulent induction at low magnetic Prandtl numbers} 

\author{A~A~Schekochihin,$^{1,2,3}$ A~B~Iskakov,$^4$ S~C~Cowley,$^{1,4}$ 
J~C~McWilliams,$^5$ M~R~E~Proctor,$^3$ T~A~Yousef\,$^3$} 

\address{$^1$ Blackett Laboratory, Imperial College, London SW7~2BW, UK}
\address{$^2$ King's College, Cambridge CB2 1ST, UK}
\address{$^3$ DAMTP, University of Cambridge, Cambridge CB3 0WA, UK}
\address{$^4$ Department of Physics and Astronomy, 
UCLA, Los Angeles, CA 90095-1547, USA}
\address{$^5$ Department of Atmospheric Sciences, 
UCLA, Los Angeles, CA 90095-1565, USA}
\ead{a.schekochihin@imperial.ac.uk}

\begin{abstract}
This paper is a detailed report on a programme of 
direct numerical simulations of incompressible nonhelical 
randomly forced MHD turbulence that are used to settle a long-standing 
issue in the turbulent dynamo theory and demonstrate 
that the fluctuation dynamo exists in the limit of large magnetic Reynolds 
number $\Rm\gg1$ and small magnetic Prandtl number $\Pm\ll1$. 
The dependence of the critical $\Rmc$ for dynamo vs.\ the 
hydrodynamic Reynolds number $\Re$ is obtained for 
$1\lesssim\Re\lesssim6700$. In the limit $\Pm\ll1$, 
$\Rmc$ is at most three times larger 
than for the previously well established 
dynamo at large and moderate Prandtl numbers: 
$\Rmc\lesssim 200$ for $\Re\gtrsim6000$ 
compared to $\Rmc\sim60$ for $\Pm\ge1$.
The stability curve $\Rmc(\Re)$ (and, it is argued, the nature 
of the dynamo) is substantially different from 
the case of the simulations and liquid-metal 
experiments with a mean flow. 
It is not as yet possible to determine numerically 
whether the growth rate of the magnetic energy 
is $\propto\Rm^{1/2}$ in the limit $\Re\gg\Rm\gg1$, 
as should be the case if the dynamo is driven 
by the inertial-range motions at the resistive scale, 
or tends to an $\Rm$-independent value comparable to 
the turnover rate of the outer-scale motions. 
The magnetic-energy spectrum in the low-$\Pm$ 
regime is qualitatively different from the $\Pm\ge1$ case 
and appears to develop a negative spectral slope, although 
current resolutions are insufficient to determine its 
asymptotic form. 
At $\Rm\in(1,\Rmc)$, the magnetic fluctuations 
induced via the tangling by turbulence of a weak mean field
are investigated and the possibility of a $k^{-1}$ spectrum 
above the resistive scale is examined. 
At low $\Rm<1$, the induced fluctuations 
are well described by the quasistatic approximation; 
the $k^{-11/3}$ spectrum is confirmed for the first time 
in direct numerical simulations. 
Applications of the results on turbulent induction 
to understanding the nonlocal energy transfer from the 
dynamo-generated magnetic field to smaller-scale magnetic 
fluctuations are discussed. 
The results reported here are of fundamental importance 
for understanding the genesis of small-scale magnetic 
field in cosmic plasmas. 
\end{abstract}
\pacs{91.25.Cw, 47.65.-d, 95.30.Qd, 96.60.Hv, 47.27.ek}
\submitto{\NJP}

\maketitle

\section{Introduction}

The dynamo effect, or amplification of magnetic field by fluid motion is a 
fundamental physical mechanism most likely to be responsible for the ubiquitous 
presence of dynamically strong magnetic fields in the Universe --- from 
planets and stars to galaxies and galaxy clusters 
\citeaffixed{Moffatt_book,Childress_Gilbert,Roberts_Glatzmaier,Ossendrijver_review,Widrow}{e.g.,}. 
It is important to distinguish between two main types of dynamo. 
The first is the large-scale, or {\em mean-field dynamo} defined as the growth of 
magnetic field at scales larger than the scale of the fluid motion. If the fluid 
is turbulent, this refers to the outer (energy-containing) scale, denoted 
here by $\lf$. In this paper, we shall not be concerned with this 
type of dynamo and concentrate on the second kind, the  
small-scale, or {\em fluctuation dynamo}, which is 
defined as the growth of magnetic-fluctuation energy 
at or below the outer scale of the motion. Note that if 
a large-scale magnetic field, dynamo-generated or otherwise, 
is present, there will always be some tangling of this field 
by the fluid, giving rise to small-scale magnetic fluctuations. 
This effect, known as the {\em magnetic induction}, should not be 
confused with the fluctuation dynamo, although it can often be 
difficult to tell which of the two is primarily 
responsible for the presence of magnetic energy at small scales. 

Despite its very generic nature, the presence of dynamo action
in any particular system is usually impossible to prove analytically. 
This is especially true in the astrophysically relevant limit of 
large hydrodynamic and magnetic Reynolds numbers, $\Re\gg1$ and $\Rm\gg1$, 
when the fluid motion is turbulent. 
Numerical experiments have, therefore, played a crucial role in 
building up the case for the turbulent dynamo 
\citeaffixed{Galloway_Proctor,Meneguzzi_Frisch_Pouquet,Cattaneo,Brandenburg}{e.g.,}. 
These have recently been joined by a successful laboratory demonstration 
of the dynamo action in a geometrically unconstrained 
turbulence of liquid sodium \cite{Monchaux_etal,Berhanu_etal}. 
However, both in the laboratory and in the computer, 
it has proven very hard to access the parameter 
regimes that are sufficiently asymptotic in both $\Re$ and $\Rm$ 
to resemble real astrophysical situations. 

The key parameter that makes the situation difficult 
is the magnetic Prandtl number, $\Pm=\Rm/\Re=\nu/\eta$ 
(viscosity/magnetic diffusivity). 
In most natural systems, $\Pm$ is either very large or very small. 
The former limit is appropriate for hot diffuse plasmas 
such as the warm and hot phases of the interstellar medium 
and the intracluster medium of galaxy clusters. 
The latter limit is realized in denser environments, e.g., 
the liquid-metal cores of planets ($\Pm\sim10^{-5}$), 
the stellar convective zones ($\Pm\sim10^{-7}-10^{-4}$ 
for the Sun, depending on the depth), and protostellar discs. 
In liquid-sodium experiments, where $\Pm\sim10^{-6}$, 
the main problem has been to access the high-$\Rm$ regime: thus, 
to get $\Rm\sim10^2$, it is neccessary to drive 
fluid flows with $\Re\sim10^8$. 

The importance of $\Pm$ lies in that it 
determines the relative size of the viscous and resistive scales 
in the system (denoted here $\lvisc$ and $\lres$, respectively). 
When $\Pm\gg1$, $\lres/\lvisc\sim\Pm^{-1/2}\ll1$ for a weak growing 
field \citeaffixed[and references therein]{SCTMM_stokes}{i.e., 
in the kinematic-dynamo regime; see}. This means that the 
resistive scale lies outside the inertial range of the turbulence, 
deep in the viscous range, where the velocity field is spatially 
smooth. In contrast, when $\Pm\ll1$ (while both $\Re\gg1$ and $\Rm\gg1$), 
one expects $\lres/\lvisc\sim\Pm^{-3/4}\gg1$ 
\cite{Moffatt}. This estimate places the magnetic cutoff in the middle 
of the inertial range, asymptotically far away both from the viscous 
and the outer scales. Thus, 
the computational challenge posed by 
large or small values of $\Pm$ is to resolve two scale separations:
$L\gg\lvisc\gg\lres$ for $\Pm\gg1$ or $L\gg\lres\gg\lvisc$ for $\Pm\ll1$. 

The fluctuation dynamo at large $\Pm$ is an easier case 
both to understand physically and to handle numerically. 
In this limit, field amplification is due to the random stretching 
of the magnetic field by the fluid motion 
\cite{Batchelor,Moffatt_Saffman,Zeldovich_etal,Childress_Gilbert,Ott_review,Chertkov_etal}. 
Since the spatially smooth viscous-scale motions have the largest 
turnover rate, they are primarily responsible for the stretching.
It is, therefore, not essential to have a large $\Re$ in order 
to capture the field growth --- all that is needed is 
a flow with chaotic trajectories, which can be spatially smooth 
(laminar). This simplification can be exploited 
to model numerically the large-$\Pm$ limit \cite{SCTMM_stokes}. 

The case of $\Pm=1$, although not encountered in nature, 
has historically been the favourite choice of convenience 
in numerical simulations 
--- it is, indeed, for this case that the fluctuation dynamo was first 
obtained in the computer \cite{Meneguzzi_Frisch_Pouquet}. 
Examination of the nature of the dynamo at $\Pm=1$ leads one 
to conclude that it belongs essentially to the 
same class as the large-$\Pm$ limit \cite{SCTMM_stokes}. 
This is because the bulk of the magnetic energy in this case 
still resides below the viscous scale. 

The physical considerations based on the random stretching
that provide a qualitative (if not mathematically rigourous)
explanation of how the fluctuation dynamo is possible 
(see references cited above) depend on the assumption 
that the scale of the fluid motion that does 
the stretching (the viscous scale $\lvisc$) is larger than 
the scale of the field that is stretched (the resistive scale $\lres$). 
This cannot be valid in the case of low $\Pm$, when $\lres\gg\lvisc$. 
Clearly, the latter limit is qualitatively different because 
the magnetic fluctuations, heavily dissipated below $\lres$, 
cannot feel the spatially smooth viscous-scale motions. 
Can the inertial-range and/or the outer-scale motions still 
make the field grow? No compelling {\em a priori} physical argument 
either for or against such a dynamo has been proposed. 
As was pointed out by \citeasnoun{Vainshtein_lowPm}, 
in the absence of a better physical understanding, 
the problem is purely quantitative: the stretching and turbulent 
diffusion effects being of the same order at each scale in 
the inertial range, one cannot predict which of them wins. 

Our first numerical investigation of the problem of low-$\Pm$ dynamo 
\cite{SCMM_lowPm} 
revealed that at fixed $\Rm$, wherever there was dynamo at $\Pm=1$, it 
weakened or disappeared if $\Re$ was increased.\footnote[1]{Previous 
attempts to simulate turbulent dynamo in various other, 
mostly convective, contexts had also found achieving a sustained 
field amplification problematic at $\Pm<1$ 
\citeaffixed{Nordlund_etal,Brandenburg_etal,Christensen_Olson_Glatzmaier}{e.g.,}.} 
Our conclusion was that, as far as we could tell at the resolutions 
available to us then, the critical magnetic Reynolds number 
$\Rm_c$ for dynamo increased with $\Re$. 
This was confirmed by \citeasnoun{Haugen_Brandenburg_Dobler}, who 
used a different (grid, rather than spectral) code. 
Our and their results, enhanced somewhat by using hyperviscosity, 
were assembled together by \citeasnoun{SHBCMM_lowPm2} to produce 
the first stability curve $\Rmc(\Re)$ for the fluctuation dynamo. 
While we were able to show that $\Rmc$ increased with $\Re$, it remained 
unknown if this increase was to be eventually saturated with $\Rmc$ reaching 
some finite limit as $\Re\to\infty$. In this paper, we give a detailed 
report on the new results that show that it does, i.e., we demonstrate 
that {\em fluctuation dynamo at low $\Pm$ exists} 
(a preliminary report appeared in \citeasnoun{ISCMP_lowPm3}). 
These results are described in \secref{sec_dynamo}. 
We also report that the form of the magnetic-energy spectrum 
changes qualitatively in the low-$\Pm$ limit (\secref{sec_spectra}). 
We further  discuss the comparison of our results with simulations 
by \citeasnoun{Ponty_etal2}, \citeasnoun{Mininni} and others 
of the fluctuation dynamo in turbulence with a mean flow (\secref{sec_meanflow}) 
and discuss the remaining theoretical uncertainties and 
unsolved questions (\secref{sec_faqs}). The most important 
of these is whether the dynamo we have found is driven by the 
inertial-range motions at the resistive scale --- if it is, 
its growth rate should be proportional to $\Rm^{1/2}$, which would make 
it a dominant (and universal) field-amplification effect. 
Finally, we proceed in \secref{sec_ind} to report 
a numerucal study of small-scale magnetic induction, an effect 
that is related rather closely to the dynamo problem. 
\Secref{sec_conc} summarizes our findings. 

\section{Fluctuation Dynamo}
\label{sec_dynamo}

\subsection{Problem Set Up}
\label{sec_setup}

We use the standard pseudospectral method to 
solve in a periodic cube the equations of incompressible MHD:
\bea
\label{u_eq}
{\dd\vu\over\dd t} + \vu\cdot\vdel\vu = -\vdel p - \nu_n|\vdel|^n\vu
+ \vB\cdot\vdel\vB + \vf,\\
\label{B_eq}
{\dd\vB\over\dd t} + \vu\cdot\vdel\vB = 
\vB\cdot\vdel\vu + \eta\nabla^2\vB.
\eea
Here $\vu$ is the velocity field, $\vB$ is the magnetic field (in velocity 
units) and the density of the fluid is taken to be constant and equal to~1. 
In order to study the field growth or decay in the kinematic regime, 
we initialize our simulations 
with random magnetic fluctuations at very small energy --- usually between 
$10^{-10}$ and $10^{-6}$ of the dynamically significant level 
--- so the Lorentz force, while retained, is never important. 
The velocity is forced by a {\em nonhelical} 
homogeneous body force, which consists 
in randomly injecting energy in a $\delta$-correlated (white-noise) fashion 
into the Fourier harmonics with wave numbers $|\vk|\le\sqrt{2}\, k_0$, 
where $k_0=2\pi$ is the box wave number (the box size is 1). 
The white-noise character of the forcing allows us to fix the 
average injected power per unit volume: $\eps=\la\vu\cdot\vf\ra=1$, 
where the angle brackets stand for volume and time averaging.

\begin{table}[t]
\caption{\label{tab_dynamo1} Index of runs --- Part I}
\lineup
{\footnotesize
\begin{tabular}{@{}lrlrllrrlll}
\br
Run    &Res.   &n&       $\nu_n\qquad$ & \Rm & $\Pm$ & $\Re$&$\Rel$&$\gamma$&$\urms$& Fig.\\
\mr
\multicolumn{11}{l}{$\eta=4\times10^{-3}$}\\
\mr
a0     & $64^3$&2&  $4\times10^{-3\ }$ &  51 & 1.0   &   51 & 34 &\-0.20  & 1.29  &\ref{fig_gamma} \z{709}\\
H01    &$128^3$&8&          $10^{-12}$ &  57 & 0.43  &  133 & 64 &\-0.145 & 1.43  &\ref{fig_gamma}, \ref{fig_spPm}(a) \z{821}\\
H02    &$128^3$&8&          $10^{-14}$ &  57 & 0.180 &  320 & 98 &\-0.89  & 1.43  &\ref{fig_gamma}, \ref{fig_spPm}(b) \z{822}\\
H03    &$128^3$&8&          $10^{-16}$ &  57 & 0.080 &  710 &148 &\-1.13  & 1.43  &\ref{fig_gamma}, \ref{fig_spPm}(d) \z{823}\\
H04    &$128^3$&8&          $10^{-17}$ &  57 & 0.052 & 1100 &184 &\-1.04  & 1.43  &\ref{fig_gamma}, \ref{fig_spPm}(e) \z{824}\\
H05    &$256^3$&8&          $10^{-18}$ &  57 & 0.034 & 1680 &230 &\-1.02  & 1.44  &\ref{fig_gamma}, \ref{fig_spPm}(f) \z{825}\\
H07    &$256^3$&8&          $10^{-20}$ &  57 & 0.0148& 3800 &342 &\-0.99  & 1.43  &\ref{fig_gamma}, \ref{fig_HM}(a) \z{804}\\
\mr
\multicolumn{11}{l}{$\eta=2\times10^{-3}$}\\
\mr
A1     &$128^3$&2&  $2\times10^{-3\ }$ & 107 & 1.0   &  107 & 53 &  0.30  & 1.35  &\ref{fig_gamma} \z{581}\\
A2     &$128^3$&2&         $10^{-3\ }$ & 110 & 0.5   &  220 & 80 &  0.139 & 1.39  &\ref{fig_gamma} \z{590}\\
A3     &$256^3$&2&  $5\times10^{-4\ }$ & 110 & 0.25  &  440 &111 &\-0.22  & 1.38  &\ref{fig_gamma} \z{633}\\
A4$^*$ &$256^3$&2&$2.5\times10^{-4\ }$ & 104 & 0.125 &  830 &148 &\-0.52  & 1.31  &\ref{fig_gamma} \zABI{$(5,20)$}\\
A5$^*$&$256^3$&2&$1.25\times10^{-4\ }$ & 106 & 0.0625& 1700 &230 &\-0.74  & 1.33  &\ref{fig_gamma} \zABI{$(5,20.8)$}\\
A6$^*$&$512^3$&2&$6.25\times10^{-5\ }$ & 105 &0.03125& 3400 &322 &\-0.41  & 1.32  &\ref{fig_gamma}, \ref{fig_spRe} \zABI{$(2,12.25)$}\\
HA1    &$128^3$&8&          $10^{-14}$ & 112 & 0.36  &  310 & 95 &  0.0106& 1.41  &\ref{fig_gamma}, \ref{fig_spPm}(a) \z{809}\\
HA2    &$128^3$&8&          $10^{-16}$ & 112 & 0.157 &  710 &144 &\-0.64  & 1.40  &\ref{fig_gamma}, \ref{fig_spPm}(b) \z{808}\\
HA3    &$128^3$&8&          $10^{-17}$ & 109 & 0.109 & 1000 &165 &\-0.60  & 1.38  &\ref{fig_gamma}, \ref{fig_spPm}(d) \z{807}\\
HA4    &$256^3$&8&          $10^{-18}$ & 114 & 0.069 & 1660 &220 &\-0.54  & 1.43  &\ref{fig_gamma}, \ref{fig_spPm}(e) \z{806}\\
HA6    &$256^3$&8&          $10^{-20}$ & 117 & 0.030 & 3900 &360 &\-0.46  & 1.47  &\ref{fig_gamma}, \ref{fig_spRe}, \ref{fig_spPm}(f), \ref{fig_HM}(a) \z{798}\\
HA8$^*$&$512^3$&8&          $10^{-22}$ & 109 & 0.0173& 6300 &420 &\-0.45  & 1.38  &\ref{fig_gamma} \zABI{$(2,13)$}\\
\mr
\multicolumn{11}{l}{$\eta=10^{-3}$}\\
\mr
B1     &$128^3$&2&         $10^{-3\ }$ & 210 & 1.0   &  210 & 80 &  0.77  & 1.34  &\ref{fig_gamma}, \ref{fig_spRm}(a) \z{582}\\
B2     &$256^3$&2&  $5\times10^{-4\ }$ & 220 & 0.5   &  440 &110 &  0.49  & 1.38  &\ref{fig_gamma}, \ref{fig_spRm}(a) \z{631}\\
B3     &$256^3$&2&$2.5\times10^{-4\ }$ & 230 & 0.25  &  900 &160 &  0.161 & 1.42  &\ref{fig_gamma}, \ref{fig_spRm}(a) \z{632}\\
B4     &$256^3$&2&$1.25\times10^{-4\ }$& 220 & 0.125 & 1760 &230 &\-0.141 & 1.38  &\ref{fig_gamma}, \ref{fig_spRm}(a) \z{635}\\
B5$^*$ &$512^3$&2&$6.25\times10^{-5\ }$& 220 & 0.0625& 3600 &330 &\-0.021 & 1.41  &\ref{fig_gamma}, \ref{fig_spRm}(a), \ref{fig_spRe}, \ref{fig_cuts_comp} \zABI{$(2,12.4)$}\\
HB1    &$128^3$&8&          $10^{-16}$ & 220 & 0.32  &  700 &143 &  0.20  & 1.41  &\ref{fig_gamma}, \ref{fig_spRm}(b), \ref{fig_spPm}(a) \z{795}\\
HB2    &$128^3$&8&          $10^{-17}$ & 230 & 0.21  & 1100 &183 &\-0.158 & 1.43  &\ref{fig_gamma}, \ref{fig_spPm}(b) \z{776}\\
HB3$^*$&$256^3$&8&          $10^{-18}$ & 220 & 0.146 & 1490 &200 &\-0.31  & 1.37  &\ref{fig_gamma}, \ref{fig_spRm}(b), \ref{fig_spPm}(c) \zABI{$(5,20.8)$}\\
HB4    &$256^3$&8&          $10^{-19}$ & 230 & 0.090 & 2500 &280 &\-0.129 & 1.44  &\ref{fig_gamma}, \ref{fig_spPm}(d) \z{830}\\
HB5    &$256^3$&8&          $10^{-20}$ & 230 & 0.057 & 4000 &360 &\-0.098 & 1.45  &\ref{fig_gamma}, \ref{fig_spRm}(b), \ref{fig_spRe}, \ref{fig_cuts_comp}, \ref{fig_spPm}(e), \ref{fig_HM}(a) \z{794}\\
HB7$^*$&$512^3$&8&          $10^{-22}$ & 220 & 0.033 & 6700 &450 &  0.089 & 1.40  &\ref{fig_gamma}, \ref{fig_spRm}(b), \ref{fig_spPm}(f) \zABI{$(2,17.6)$}\\
\br
\end{tabular}
}
\end{table}

\begin{table}[t]
\caption{\label{tab_dynamo2} Index of runs --- Part II}
\lineup
{\footnotesize
\begin{tabular}{@{}lrlrllrrlll}
\br
Run    &Res.   &n&       $\nu_n\qquad$ & \Rm & $\Pm$ & $\Re$&$\Rel$&$\gamma$&$\urms$& Fig.\\
\mr
\multicolumn{11}{l}{$\eta=7.5\times10^{-4}$}\\
\mr
HX1    &$256^3$&8&         $10^{-17}$ & 300 & 0.28  & 1050 &175 &  0.178 & 1.41  &\ref{fig_gamma}(b), \ref{fig_spPm}(a) \z{836}\\
HX2    &$256^3$&8&         $10^{-18}$ & 310 & 0.176 & 1760 &240 &\-0.033 & 1.46  &\ref{fig_gamma}(b), \ref{fig_spPm}(b) \z{833}\\
HX4    &$256^3$&8&         $10^{-20}$ & 310 & 0.087 & 3600 &340 &\-0.054 & 1.46  &\ref{fig_gamma}(b), \ref{fig_spPm}(d) \z{827}\\
\mr
\multicolumn{11}{l}{$\eta=5\times10^{-4}$}\\
\mr
C1     &$256^3$&2&  $5\times10^{-4\ }$ & 440 & 1.0   & 440  &111 &  1.40  & 1.39  &\ref{fig_gamma}, \ref{fig_cuts}, \ref{fig_spRm}(c) \z{583}\\
C2     &$256^3$&2&$2.5\times10^{-4\ }$ & 430 & 0.5   & 870  &159 &  0.89  & 1.36  &\ref{fig_gamma}, \ref{fig_spRm}(c) \z{699}\\
C3     &$256^3$&2&$1.25\times10^{-4\ }$& 440 & 0.25  & 1760 &230 &  0.49  & 1.38  &\ref{fig_gamma}, \ref{fig_spRm}(c) \z{700}\\
C4$^*$ &$512^3$&2&$6.25\times10^{-5\ }$& 450 & 0.125 & 3600 &360 &  0.21  & 1.42  &\ref{fig_gamma}, \ref{fig_spRm}(c), \ref{fig_spRe} \zABI{(2,15)}\\
HC1    &$256^3$&8&          $10^{-18}$ & 450 & 0.28  & 1620 &220 &  0.43  & 1.41  &\ref{fig_gamma}, \ref{fig_spRm}(d), \ref{fig_spPm}(a) \z{793}\\
HC2    &$256^3$&8&          $10^{-19}$ & 450 & 0.183 & 2500 &270 &  0.20  & 1.42  &\ref{fig_gamma}, \ref{fig_spPm}(b) \z{805}\\
HC3    &$256^3$&8&          $10^{-20}$ & 470 & 0.110 & 4300 &380 &  0.120 & 1.48  &\ref{fig_gamma}, \ref{fig_spRm}(d), \ref{fig_spRe}, \ref{fig_spPm}(c) \z{792}\\
HC4$^*$&$512^3$&8&          $10^{-21}$ & 460 & 0.090 & 5100 &400 &  0.27  & 1.44  &\ref{fig_gamma}, \ref{fig_spPm}(d) \zABI{$(2,17.4)$}\\
HC5$^*$&$512^3$&8&          $10^{-22}$ & 430 & 0.070 & 6200 &410 &  0.26  & 1.36  &\ref{fig_gamma}, \ref{fig_cuts}, \ref{fig_spRm}(d), \ref{fig_spPm}(e) \zABI{$(2,12.4)$}\\
\mr
\multicolumn{11}{l}{$\eta=2.5\times10^{-4}$}\\
\mr
D1$^*$ &$512^3$&2&$2.5\times10^{-4\ }$ & 810& 1.0   &  810 &141 &  1.85  & 1.27 &\ref{fig_gamma} \zABI{$(2,9)$}\\
D2$^*$ &$512^3$&2&$1.25\times10^{-4\ }$& 830& 0.5   & 1660 &210 &  1.41  & 1.31 &\ref{fig_gamma} \zABI{$(2,10)$}\\
D3$^*$ &$512^3$&2&$6.25\times10^{-5\ }$& 830& 0.25  & 3300 &320 &  0.91  & 1.31 &\ref{fig_gamma}, \ref{fig_spRe} \zABI{$(2,12)$}\\
HD1$^*$&$512^3$&8&          $10^{-20}$ & 900& 0.24  & 3800 &340 &  0.80  & 1.42 &\ref{fig_gamma}, \ref{fig_spRe}, \ref{fig_spPm}(a) \zABI{$(2,6.2)$}\\
HD3$^*$&$512^3$&8&          $10^{-22}$ & 850& 0.145 & 5900 &390 &  0.59  & 1.34 &\ref{fig_gamma}, \ref{fig_spPm}(c) \zABI{$(2,12.8)$}\\
\br
\end{tabular}
}
\end{table}

In order to increase the range of Reynolds numbers amenable to computation 
at a finite resolution, we make use both of the Laplacian viscosity 
($n=2$ in \eqref{u_eq}) and of the 8th-order hyperviscosity ($n=8$). 
For the hyperviscous runs, we define the effective viscosity 
\bea
\label{nueff_def}
\nueff = {\eps\over\la|\vdel\vu|^2\ra}.
\eea
Using the hyperviscosity appears to be a sensible way of treating the 
$\Pm\ll1$ limit because in this limit, the magnetic cutoff scale is 
much larger than the (hyper)viscous scale, $\lres\gg\lvisc$, and 
the magnetic properties of the system should not depend on the 
particular form of the viscous cutoff. In our earlier work 
\cite{SHBCMM_lowPm2}, it was confirmed that,  
as far as the values of $\Rmc$ are concerned, 
this is approximately true for a number of different choices of 
the viscous regularization, including the 4th-, 6th, and 8th-order 
hyperviscosity and the Smagorinsky LES viscosity 
(see further discussion in \secref{sec_univ}). 
It is the effective viscosity given by \eqref{nueff_def} that is 
used for calculating $\Re$ and $\Pm$ in the hyperviscous runs. 
Thus, we define 
\bea
\label{Re_def}
\fl
\Rm = {\sqrt{\usq}\over\eta k_0},\quad
\Pm = {\nu\over\eta},\quad
\Re = {\sqrt{\usq}\over\nu k_0},\quad
\Rel = {1\over\nu}\sqrt{{5\over3}{\usq^2\over\la|\vdel\vu|^2\ra}},
\eea
where $\nu=\nu_2$ for the Laplacian runs and $\nu=\nueff$ for 
the hyperviscous ones. 
We follow an established convention by using 
the box wave number $k_0=2\pi$ in the definitions of 
$\Rm$ and $\Re$. 
A characteristic of turbulence independent of this choice is 
$\Rel$, the Reynolds number based on the Taylor microscale
(also defined above).

The maximum resolution that we could afford was $512^3$. 
All our runs are summarized in \tabref{tab_dynamo1} 
and \tabref{tab_dynamo2}, where we give for each run its resolution, 
the values of the (hyper)viscosity $\nu_n$, $\Rm$, $\Pm$, $\Re$, $\Rel$, 
the growth/decay rate $\gamma$, the rms velocity $\urms=\sqrt{\usq}$, 
and a list of figures in which this run appears. 
The runs marked with a star were done using a new code written by 
A.~Iskakov. All other runs were done using another code, 
written by J.~Maron, which was the code used in our earlier 
papers \cite{SCTMM_stokes,SCMM_lowPm,SHBCMM_lowPm2}. 
Both codes are pseudospectral, solve the same equations (\eqsand{u_eq}{B_eq})
and use the same units of length 
and time, but different time-stepping, FFT and parallelisation 
algorithms, as well as slightly different implementations of the 
random forcing. We have checked conclusively that A.~Iskakov's 
code correctly reproduces the older results obtained with 
J.~Maron's code. 

\begin{figure}[t]
\begin{center}
\begin{tabular}{ccc}
\psfig{file=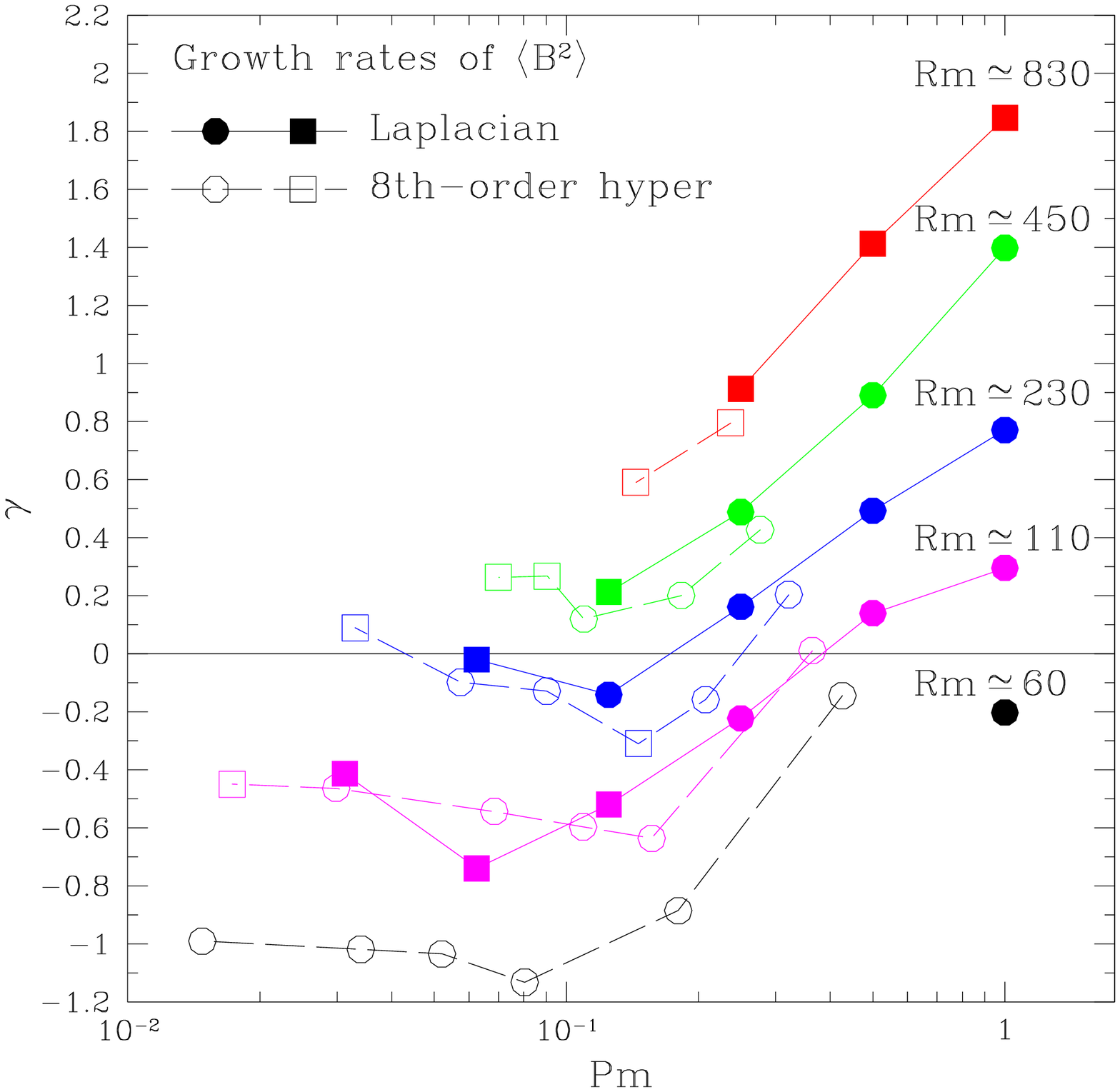,width=7.5cm}&&\psfig{file=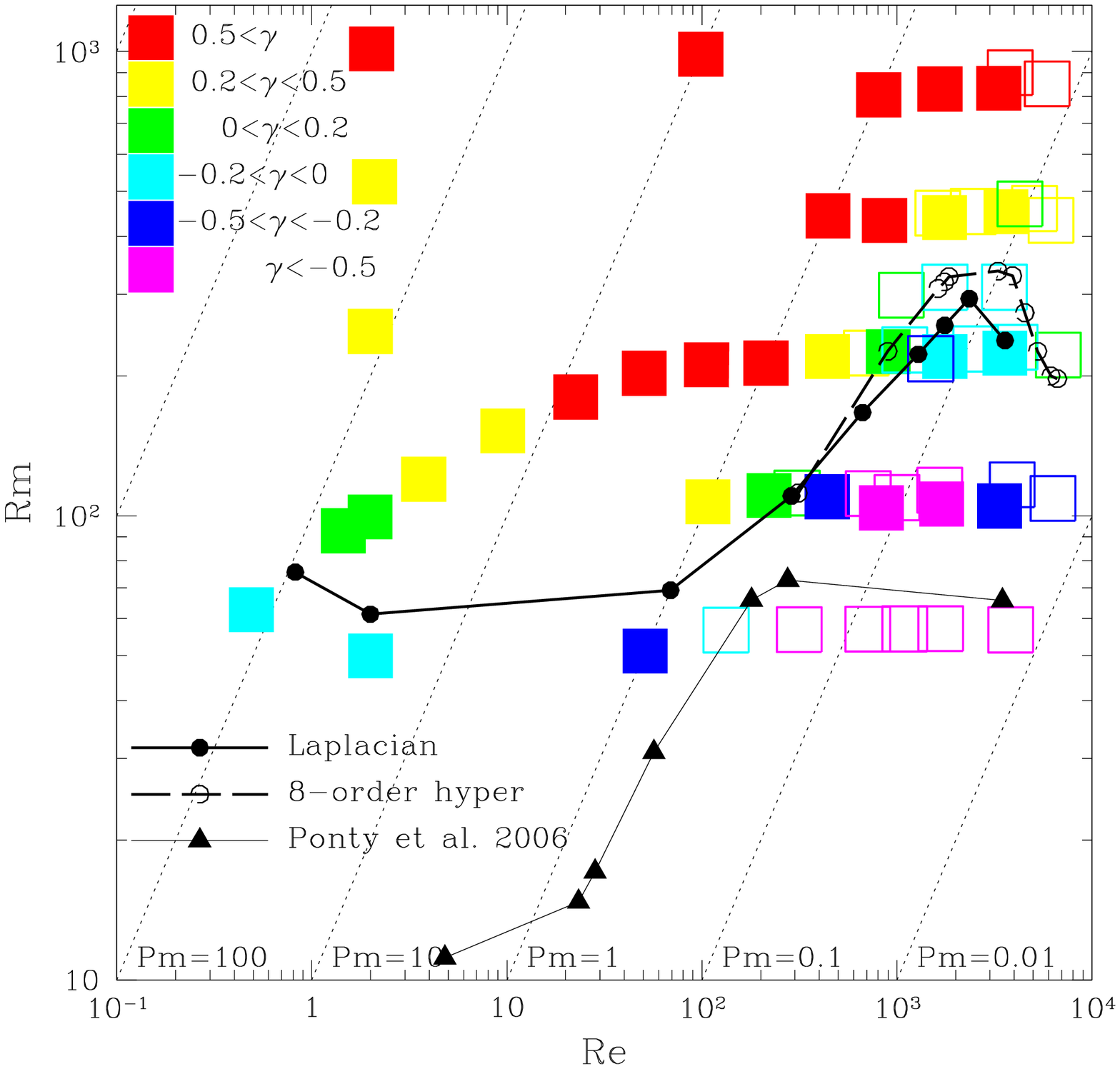,width=7.5cm}\\ 
(a) && (b)
\end{tabular}
\end{center}
\caption{\label{fig_gamma} (a) Growth/decay rate of $\Bsq$ vs.~$\Pm$ for 
five values of $\Rm$ ($\Rm\sim60$: run a0 and series H0; 
$\Rm\sim110$: series A and HA ; 
$\Rm\sim230$: series B and HB; 
$\Rm\sim450$: series C and HC;
$\Rm\sim830$: series D and HD). 
The round points were obtained using the code written by 
J.~Maron, the square points using the code written by A.~Iskakov. 
The unit of the growth rate is approximately one inverse 
turnover time at the outer scale (the precise units of time 
are set by $\eps=1$ and the box size~$=1$.)
(b) Growth/decay rates in the parameter space 
$(\Re,\Rm)$. The filled points correspond to runs with Laplacian viscosity, 
the empty ones to runs with 8th-order hyperviscosity. 
The interpolated stability curves $\Rmc(\Re)$ 
based on the Laplacian and hyperviscous runs are shown separately. 
For comparison, we also plot the $\Rmc(\Re)$ curve 
obtained by \citeasnoun{Ponty_etal2} for the turbulence with 
a mean flow (see \secref{sec_meanflow} for discussion).} 
\end{figure}

\begin{figure}[t]
\begin{center}
\begin{tabular}{ccc}
\psfig{file=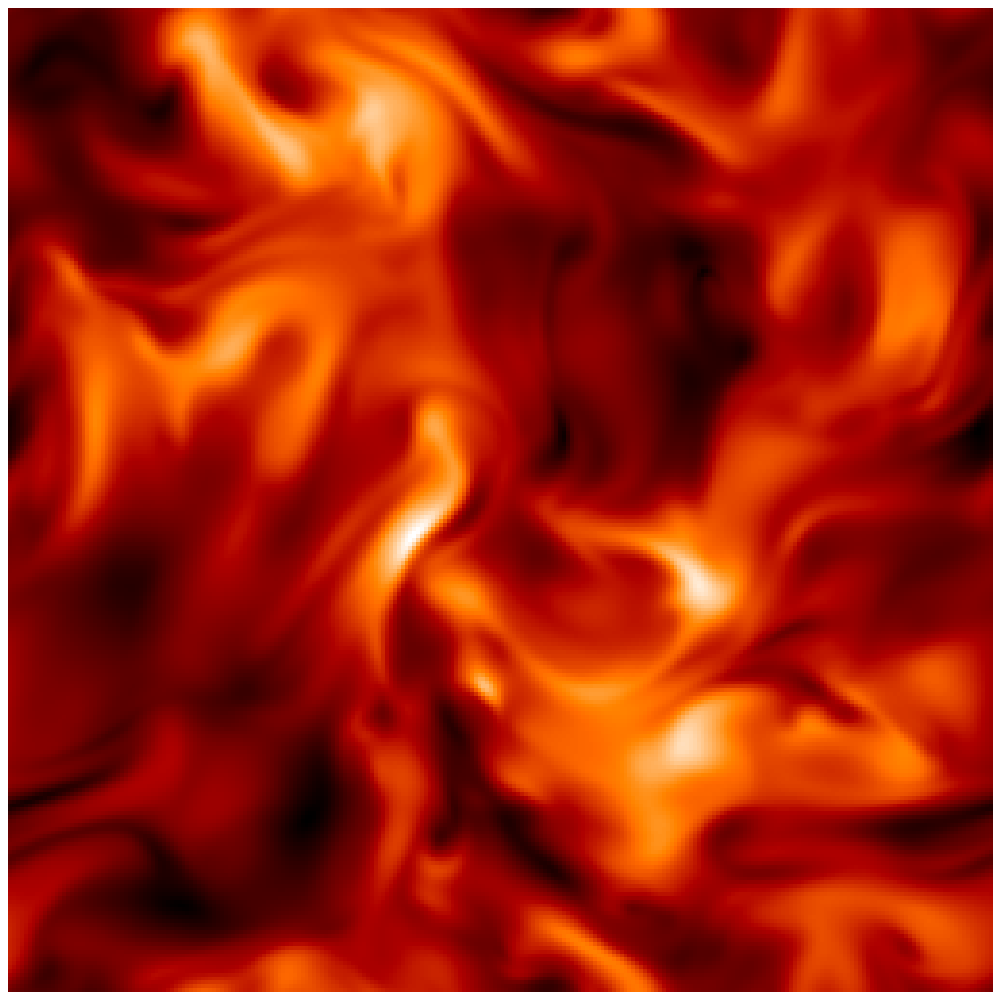,width=7.25cm}  && \psfig{file=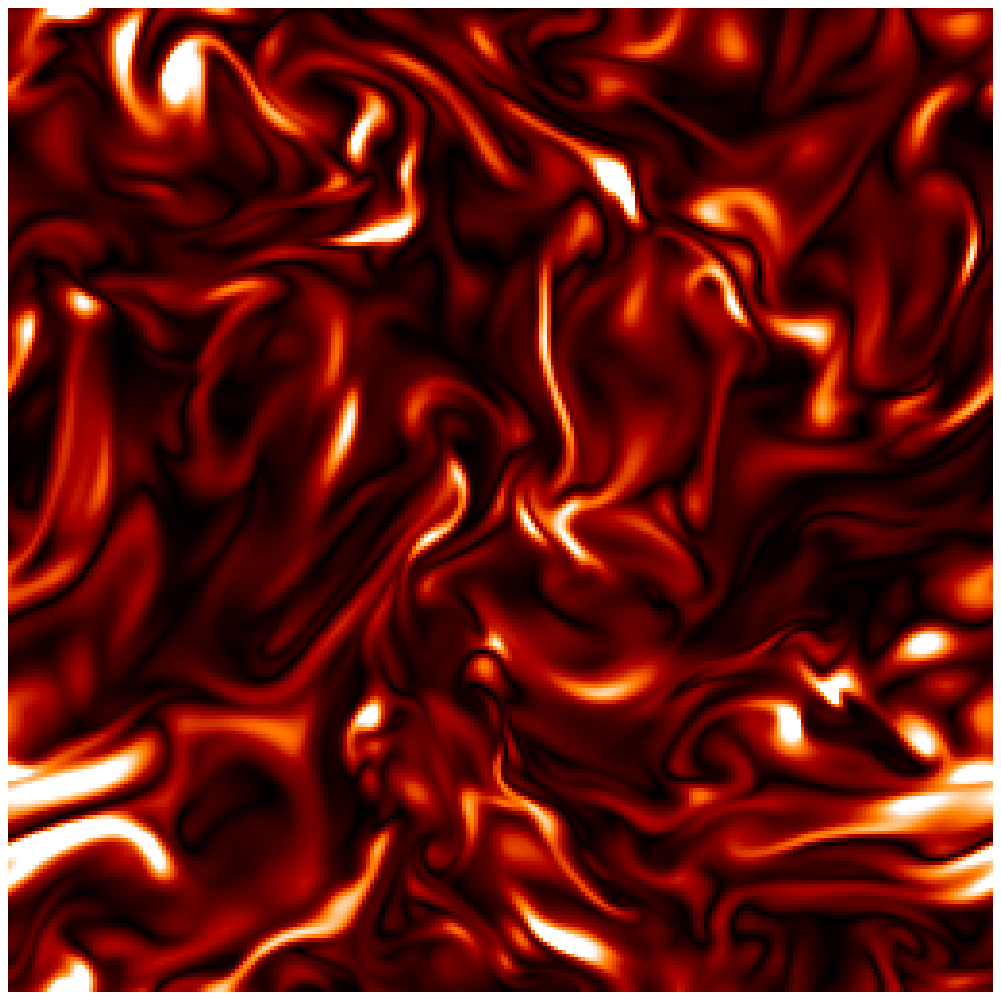,width=7.25cm}\\\\
\psfig{file=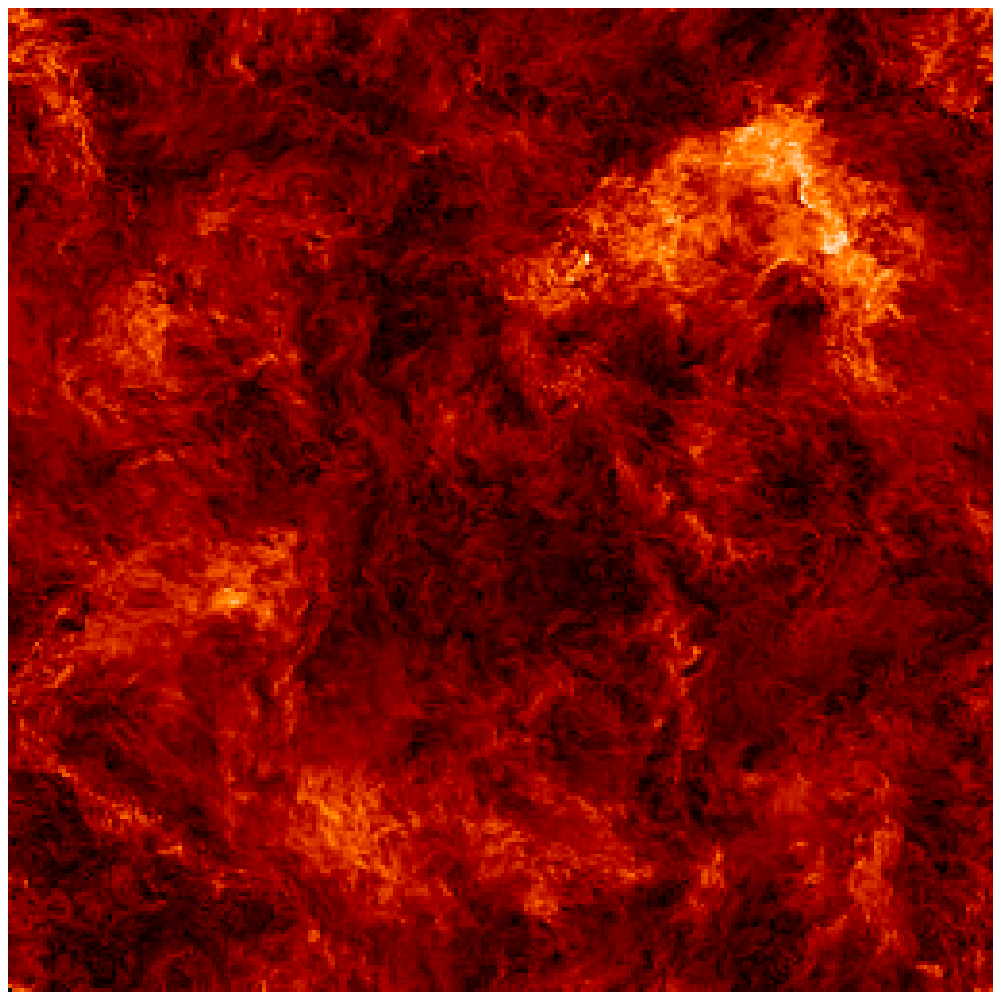,width=7.25cm} && \psfig{file=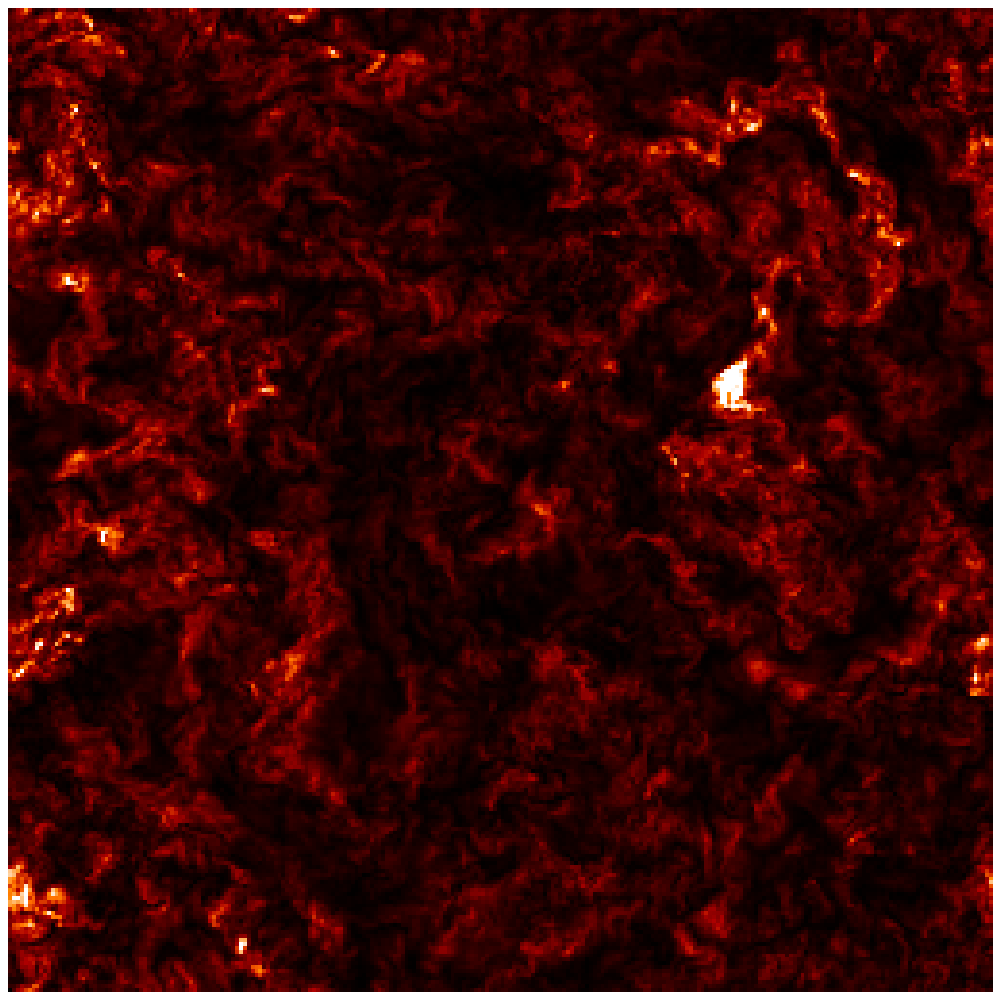,width=7.25cm}
\end{tabular}
\end{center}
\caption{\label{fig_cuts} Cross-sections of the absolute 
values of the velocity (left) and the (growing) magnetic field (right) 
for two runs with similar $\Rm$: run C1, $\Rm=\Re\simeq440$ (top) 
and run HC5, $\Rm\simeq430$, $\Re\simeq6200$ (bottom). The 
lighter/darker regions correspond to stronger/weaker fields.}
\end{figure}

\subsection{Results: Existence of the Dynamo}
\label{sec_Rmc}

In \figref{fig_gamma}(a), we show the growth/decay rates 
of the magnetic energy $\Bsq$ vs.\ $\Pm$ 
for five sequences of runs. The runs in each sequence have 
the same fixed value of $\eta$ 
and, consequently, approximately the same value of $\Rm$.  
Thus, decreasing $\Pm$ in each of these sequences is achieved 
by increasing $\Re$ while keeping $\Rm$ fixed. 
This is the same strategy as was employed by \citeasnoun{SCMM_lowPm}. 
\Figref{fig_gamma}(a) extends their figure 1(b), 
which showed runs A1--A3 and B1--B4. 

The growth rates are 
calculated via a least-squares fit to the evolution 
of $\ln\lt(\Bsq\rt)$ vs.~time. The run times were chosen so that a 
converged value of the growth/decay rate could be obtained. 
In most cases, this requires no more than 10 to 20 box-crossing times, 
although for near-marginal cases ($|\gamma|\lesssim0.1$; see 
\tabref{tab_dynamo1} and \tabref{tab_dynamo2}), 
the convergence is quite poor. This is because close to criticality, 
the evolution of the total magnetic energy has very long 
time correlations, with long periods of virtually zero change 
punctuated by periods of 
growth or decay. Fitting such a time evolution to a single 
growth/decay rate is not a particularly precise operation. 

The trend manifested in \figref{fig_gamma}(a) 
is clear: as $\Pm$ is decreased, the growth rate decreases, 
passes through a minimum and then saturates at a constant 
value, i.e., 
\bea
\label{gamma_sat}
\gamma(\Rm,\Re)\to
\gammainf(\Rm)=\const~{\rm as}~\Re\to\infty~{\rm and}~\Rm~{\rm is~fixed.}
\eea 
That such a limit should exist  
is natural because as $\Re\to\infty$ at fixed 
$\Rm$, we have $\lres\gg\lvisc\sim\lf\Re^{-3/4}\to0$ 
and one cannot expect the magnetic field to ``know'' 
exactly how small the viscous cutoff scale is. 
A much more significant result is that as $\Rm$ increases, 
the entire curve $\gamma(\Re,\Rm)$ is lifted upwards, 
so both the minimum and the asymptotic value of $\gamma$ 
are positive for $\Rm\sim450$ and above. For $\Rm\sim230$, 
$\gamma$ becomes negative approximately at $\Pm\sim0.2$
but the curve crosses zero again at $\Pm\sim0.03$ and 
emerges on the positive side, so the asymptotic 
value is expected to be positive. 
While we are unable at current resolutions to obtain the asymptotic 
values of $\gamma$ for the growing cases, we consider 
the evidence presented in \figref{fig_gamma}(a) sufficient 
to claim that such values exist and are positive. 

Thus, the fluctuation dynamo does exist in the 
nonhelical turbulence of conducting fluid with low $\Pm$. 
It is perhaps interesting to have a glimpse of what this turbulence 
``looks like.'' Snapshots of the velocity and the growing magnetic field 
for a run (HC5) with $\Pm\simeq0.07$ and $\Rm\simeq430$ 
are shown in \figref{fig_cuts} and contrasted with similar 
snapshots for a run (C1) that has approximately the same 
value of $\Rm\simeq440$ 
but $\Pm=1$ and in which the magnetic field 
exhibits a folded structure characteristic of the fluctuation 
dynamo at $\Pm\ge1$ \cite{SCTMM_stokes,Brandenburg_Subramanian,Wilkin_Barenghi_Shukurov}. 

\Figref{fig_gamma}(b) presents the magnetic-energy growth/decay rates 
in the two-dimensional parameter space $(\Re,\Rm)$. 
We also include the growth/decay rates for the $\Pm\ge1$ 
runs published in \citeasnoun{SCMM_lowPm} and 
\citeasnoun{SCTMM_stokes} to give a complete picture of what 
is known about the dependence $\gamma(\Re,\Rm)$ 
(these runs are not shown in \tabref{tab_dynamo1} and \tabref{tab_dynamo2}). 
We are now able to reconstruct the stability curve $\Rmc(\Re)$: 
each point on the curve is a linear interpolation between a decaying 
and a growing case (this is done separately for the Laplacian 
and hyperviscous runs). We see that $\Rmc$ increases with $\Re$, 
reaches a maximum around $\Rmcmax\sim350$ and $\Re\sim3000$, 
and then decreases (rather sharply). 
We expect that 
\bea
\Rmc(\Re)\to\Rmcinf=\const~{\rm as}~\Re\to\infty,
\eea
again on the grounds that exactly where the viscous cutoff is 
cannot matter in this limit, but we cannot as yet obtain 
the asymptotic value $\Rmcinf$. Discounting the unlikely possibility 
that the stability curve has multiple extrema at larger $\Re$, 
we expect the asymptotic value to be $\Rmcinf\lesssim200$. 
Above this value of $\Rm$, there is always dynamo action at small 
enough $\Pm$. Note that this represents an increase of only about 
a factor of 3 compared to the well-known critical value $\Rmc\simeq60$ 
for the fluctuation dynamo at $\Pm\ge1$ 
(established first by \citeasnoun{Meneguzzi_Frisch_Pouquet}
and confirmed in many subsequent numerical studies, e.g.,  
\citeasnoun{SCTMM_stokes}, \citeasnoun{Haugen_Brandenburg_Dobler}). 

\begin{figure}[t]
\begin{tabular}{ccc}
\psfig{file=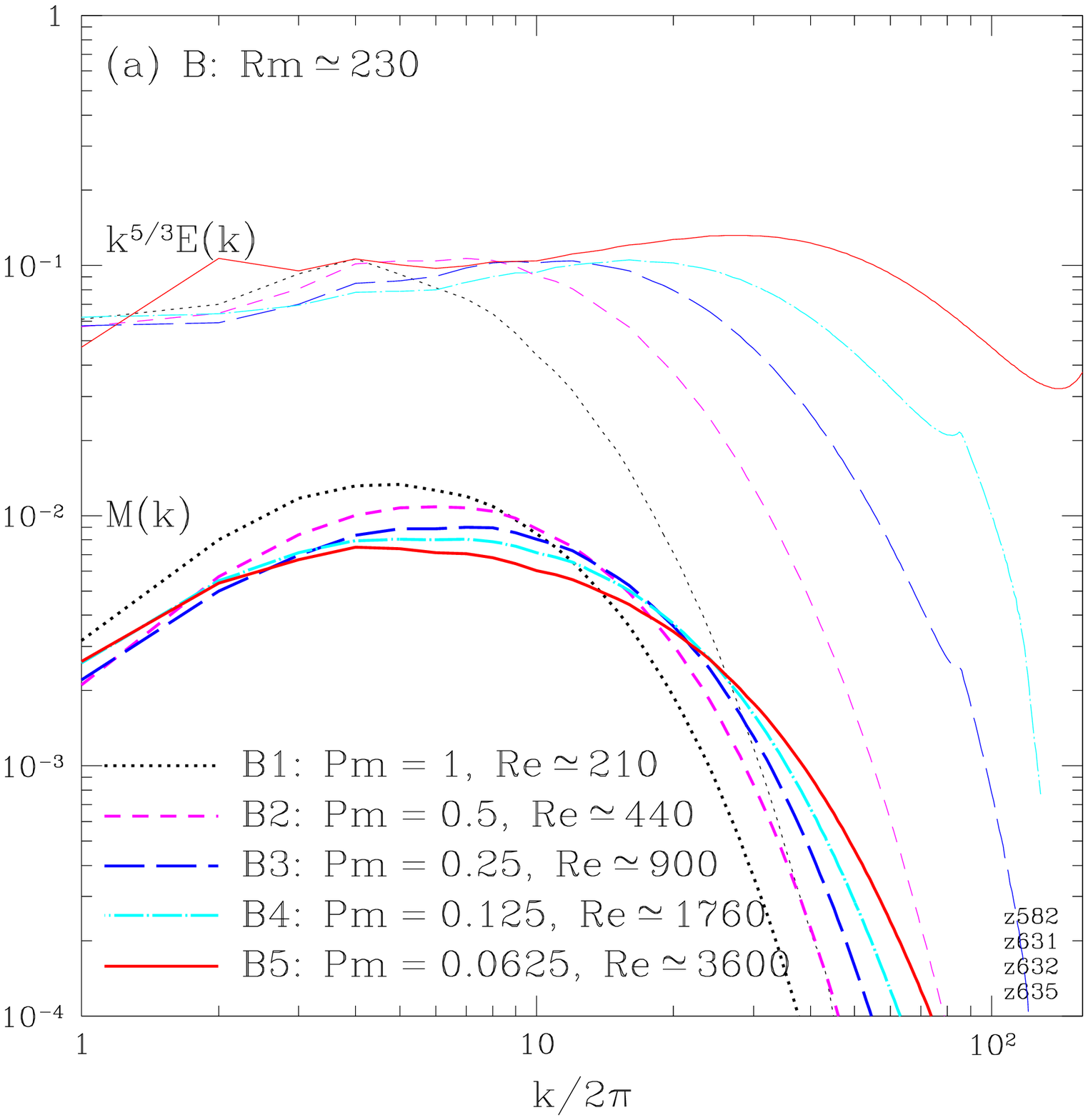,width=7.5cm} && \psfig{file=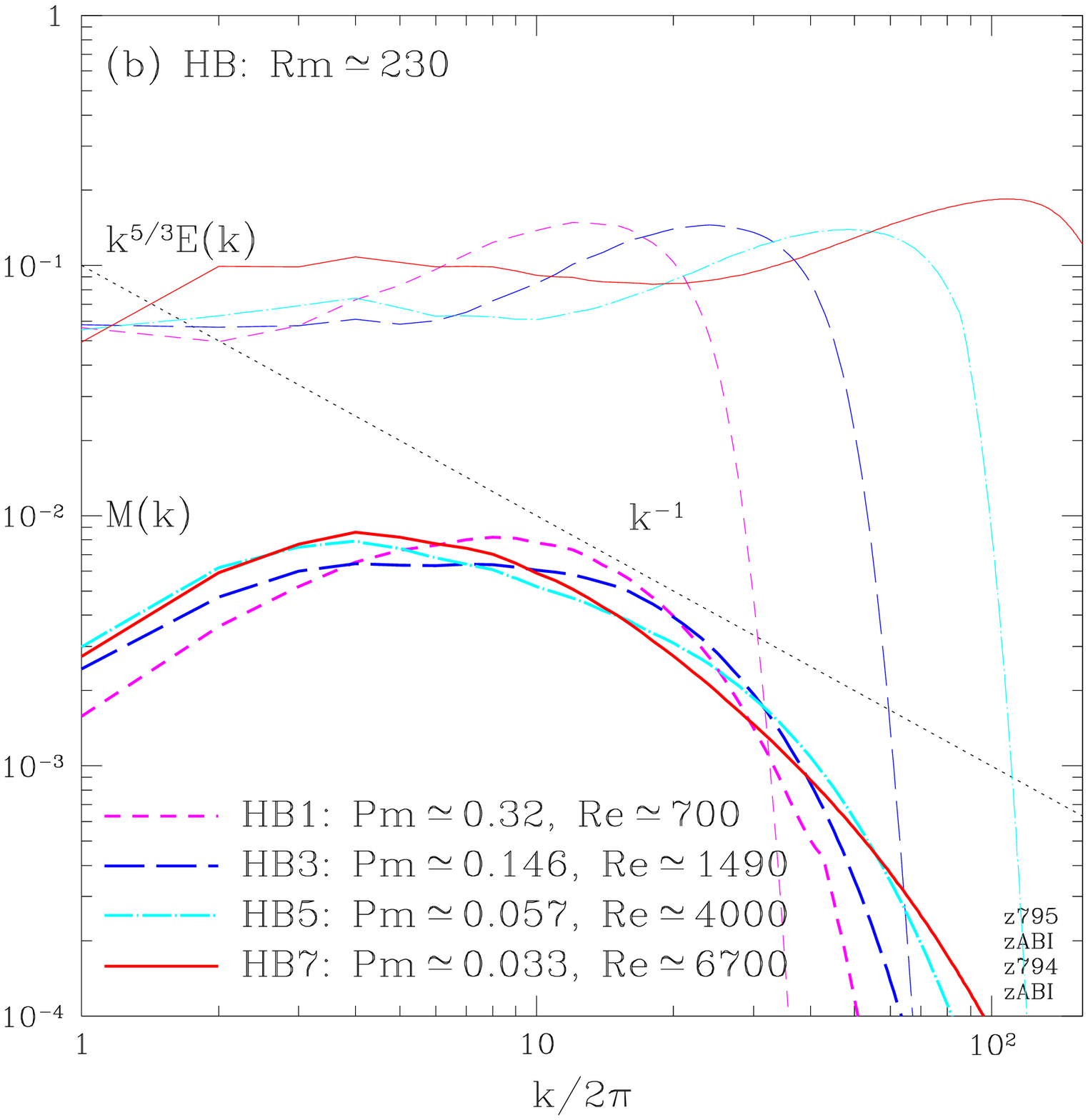,width=7.5cm}\\ 
\psfig{file=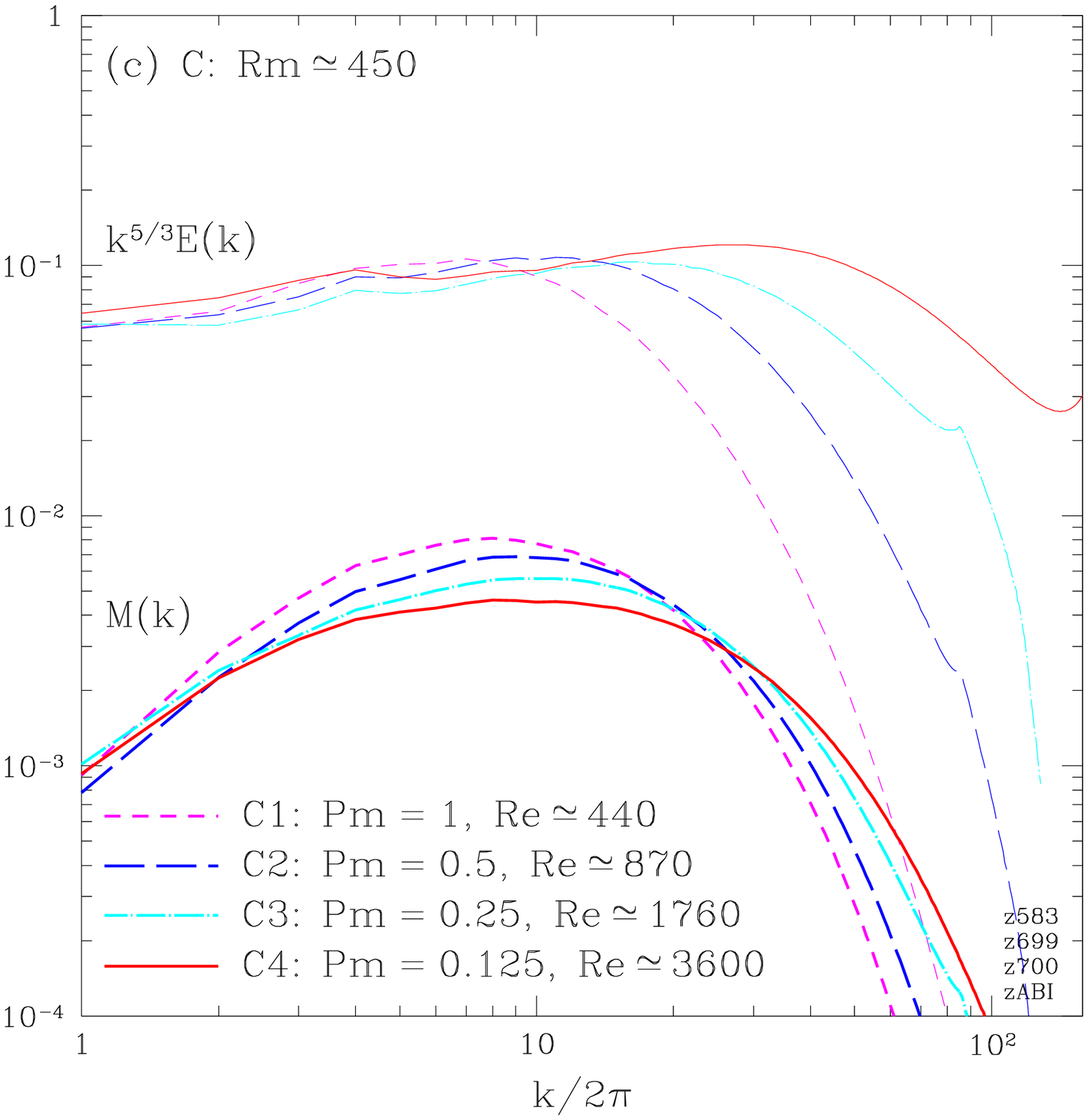,width=7.5cm} && \psfig{file=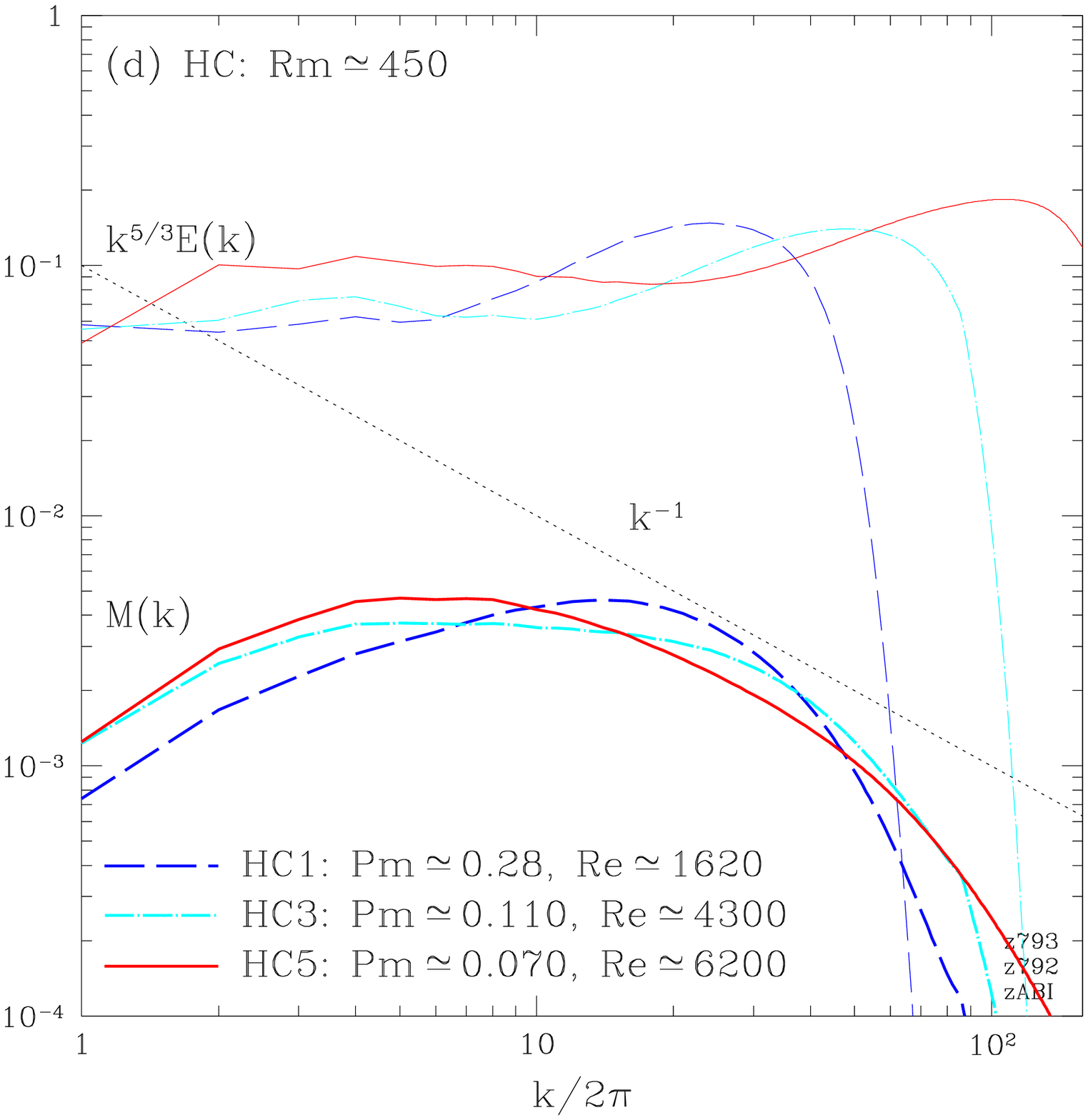,width=7.5cm}
\end{tabular}
\caption{\label{fig_spRm} Spectra of the kinetic energy 
(normalized by $\usq/2$, time-averaged and compensated by $k^{5/3}$) 
and of the magnetic energy 
(normalized at each time by the istantaneous value of 
$\Bsq/2$ and time-averaged) 
for the kinematic decaying/growing cases with 
fixed $\Rm$ and increasing $\Re$: 
(a) B series (Laplacian, $\Rm\sim230$), 
(b) HB series (hyperviscous, $\Rm\sim230$), 
(c) C series (Laplacian, $\Rm\sim450$), 
(d) HC series (hyperviscous, $\Rm\sim450$). 
In all of these cases, $\Bsq\ll\usq$. 
The $k^{-1}$ slope is given for reference and discussed in \secref{sec_highRm}.} 
\end{figure}

\Figref{fig_gamma}(b) extends figure 2 of \citeasnoun{SHBCMM_lowPm2}, 
who could only see the increasing part of the curve. They 
also reported the $\Rmc(\Re)$ dependence obtained from 
the Laplacian, 6th-order hyperviscous and Smagorinski-LES 
simulations using the grid-based PENCIL code. 
While the low-$\Pm$ fluctuation dynamo has yet 
to be found using this code, a comparison with \figref{fig_gamma}(b) 
shows that the (presumed) maximum of the stability curve for the PENCIL-code 
dynamo should lie significantly above the value we have found in our 
simulations: apparently at $\Rmcmax\gtrsim500$ and 
$\Re\gtrsim3000$. This raises the question of how universal 
the results we are reporting are. Another piece of numerical 
evidence that leads to the same question is the 
(small but measurable) difference between the stability curves 
reconstructed using the Laplacian and the hyperviscous runs. 

\begin{figure}[t]
\centerline{\psfig{file=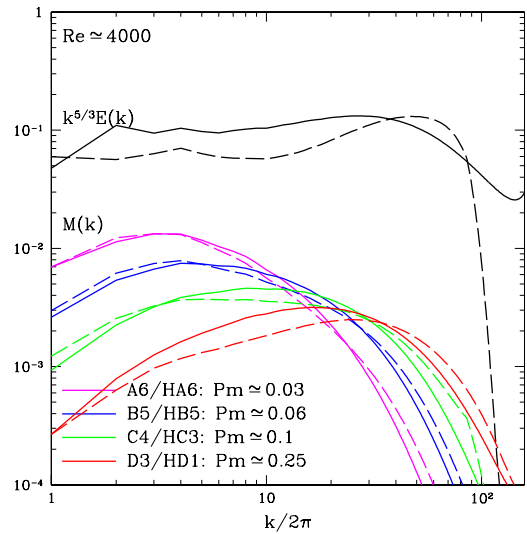,width=7.5cm}}
\caption{\label{fig_spRe} Spectra of the kinetic and magnetic 
energies (normalized as in \figref{fig_spRm}) 
for the kinematic decaying/growing cases with 
fixed $\Re\simeq4000$ and different $\Pm$. 
For each value of $\Rm$, we plot the spectra for 
a Laplacian (solid lines) and a hyperviscous 
(dashed lines) run. The kinetic-energy spectra 
are independent of $\Rm$.} 
\end{figure}

\begin{figure}
\begin{center}
\begin{tabular}{ccc}
\psfig{file=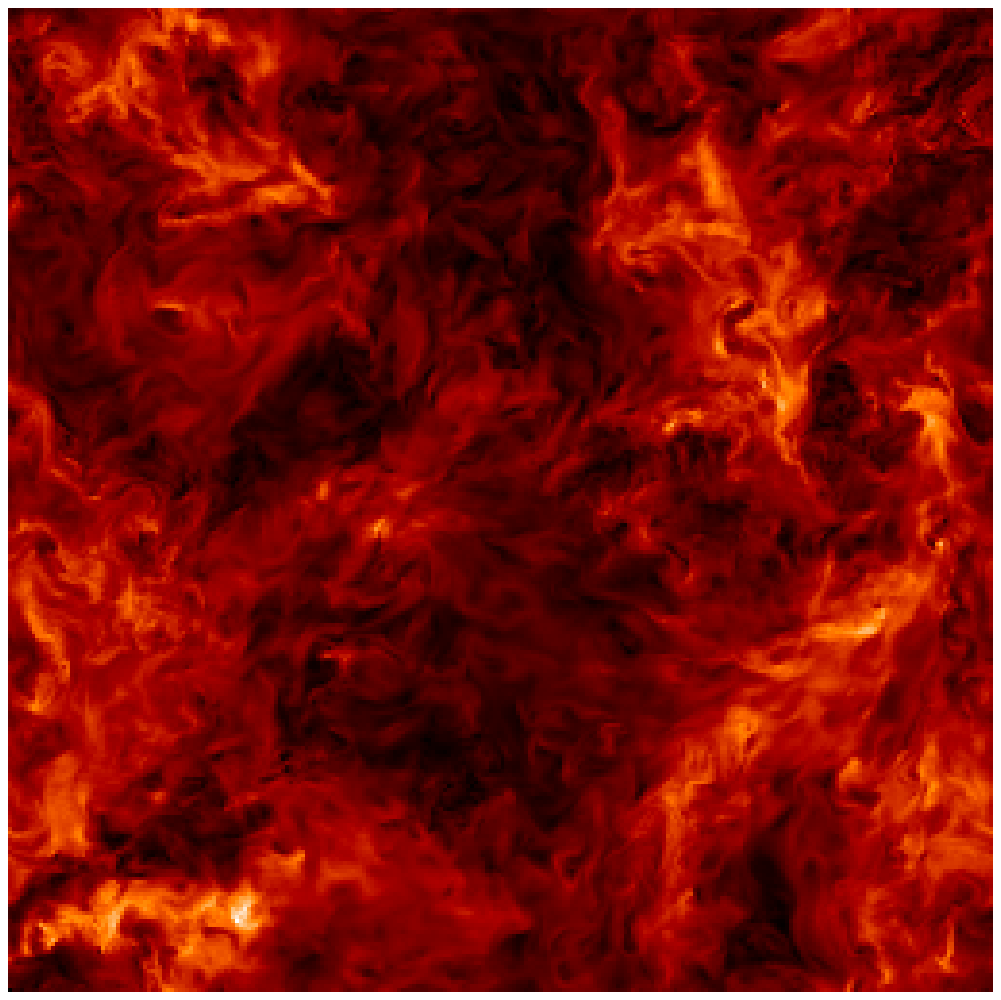,width=7.25cm}  && \psfig{file=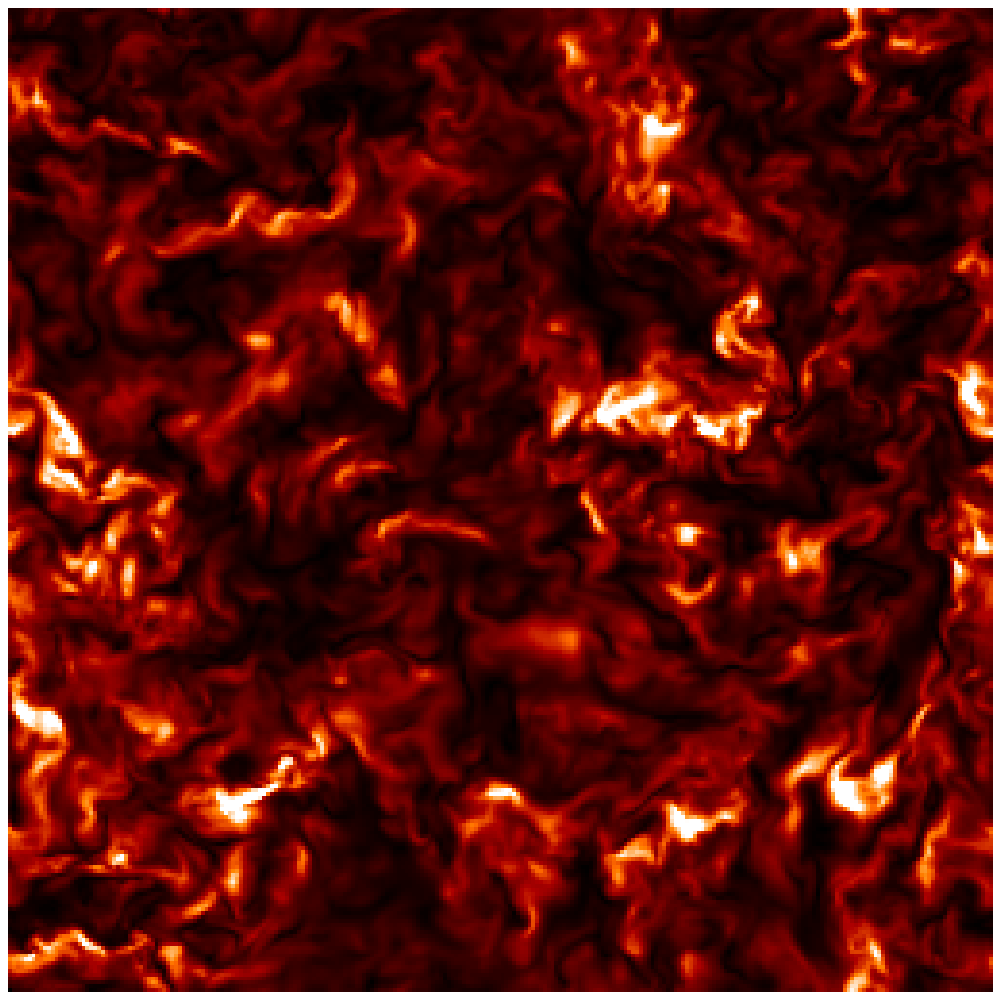,width=7.25cm}\\\\
\psfig{file=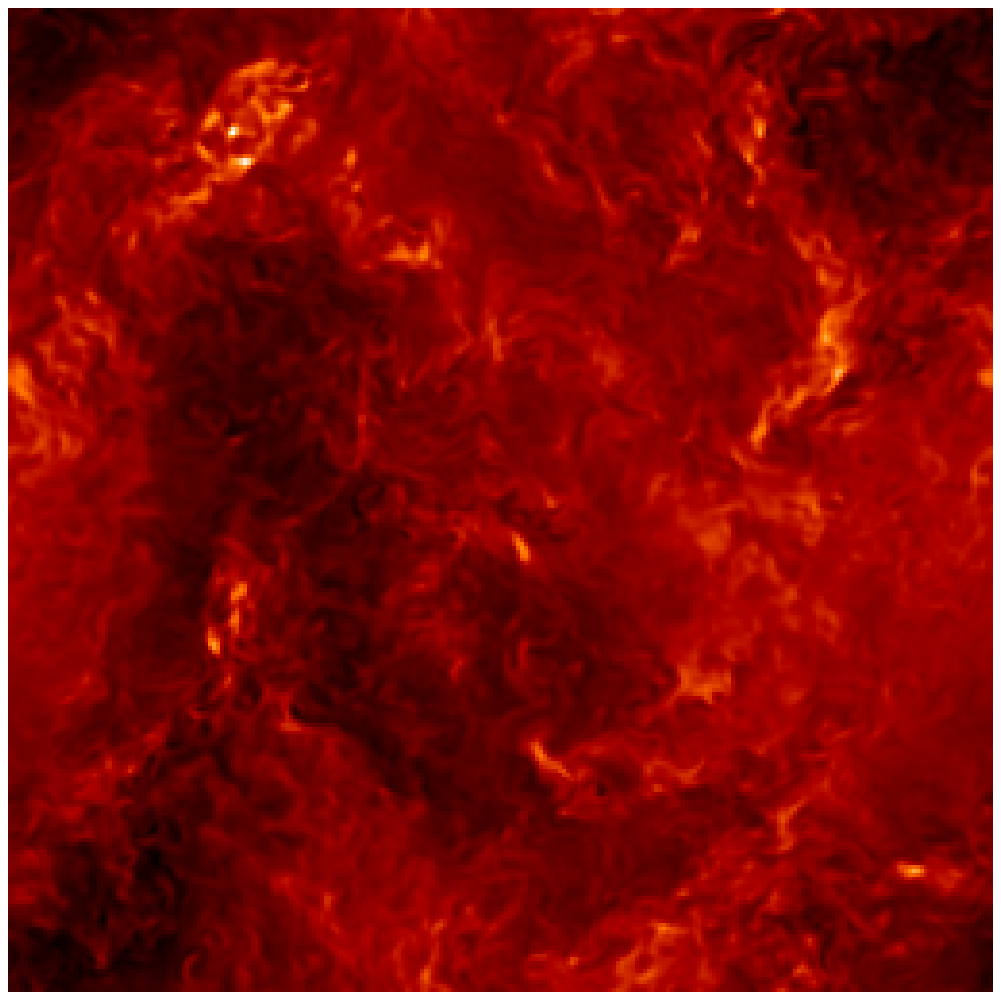,width=7.25cm} && \psfig{file=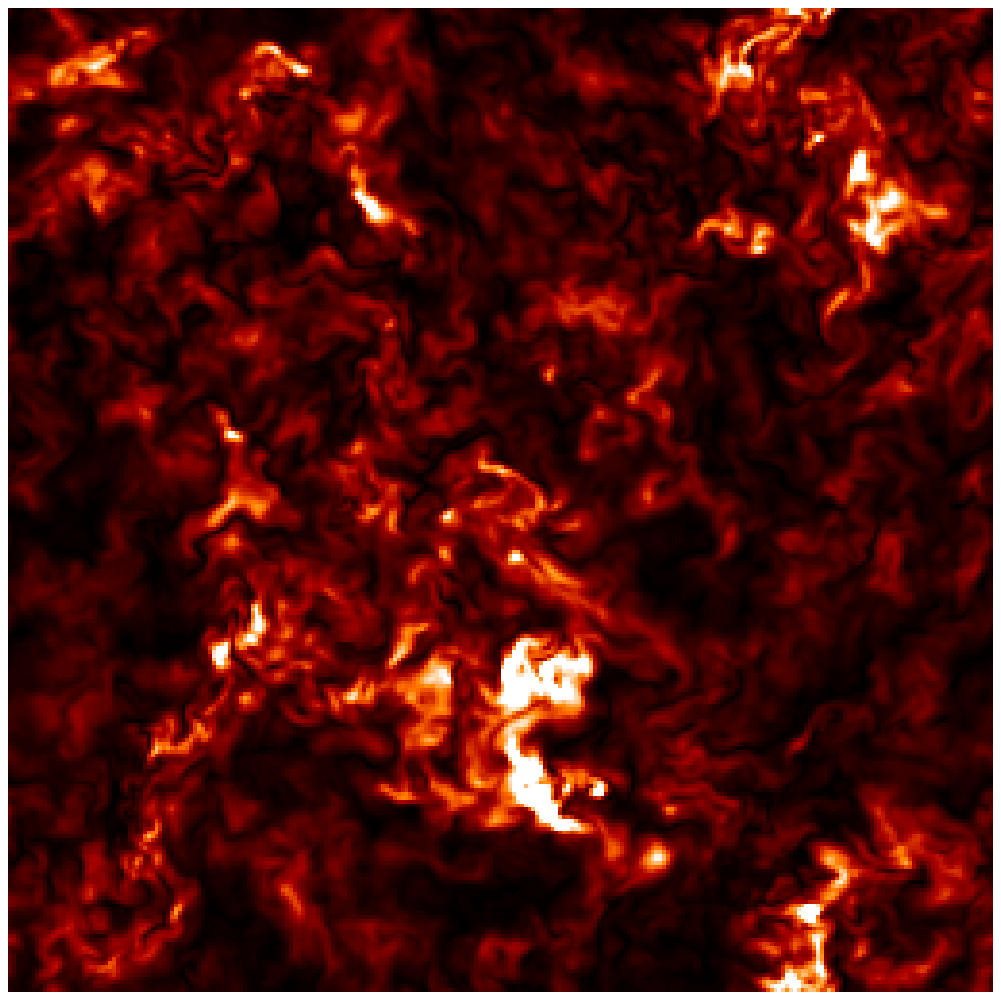,width=7.25cm}
\end{tabular}
\end{center}
\caption{\label{fig_cuts_comp} Cross-sections of the absolute 
values of the velocity (left) and the magnetic field (right) 
for a Laplacian run (B5, top) and a hyperviscous run (HB5, bottom), 
both with $\Rm\sim200$ and $\Re\sim4000$. Note that these runs 
are near-marginal with respect to the field growth (see \tabref{tab_dynamo1}
and \figref{fig_gamma}(a)).} 
\end{figure}

\subsection{Laplacian vs.\ Hyperviscous Simulations}
\label{sec_univ}

As we argued in the Introduction, it is reasonable to assume 
that whether or not the dynamo action is present in the limit 
of $\Rm\ll\Re\to\infty$ does not depend on the nature of 
the viscous cutoff. If this is true, demonstrating that it is 
present in the hyperviscous case should be sufficient to claim 
that it is present generally. It is also likely that 
the asymptotic value $\Rmcinf$ will prove to be robust. 
However, the {\em shape} of the stability curve 
$\Rmc(\Re)$ is certainly not universal. Indeed, let us consider 
what determines this shape in the transition region 
between $\Rmc\simeq60$ for $\Pm\ge1$ and the yet undetermined 
asymptotic value $\Rmcinf\lesssim200$ for $\Pm\ll1$. 
When $\Re\le\Rm$ ($\Pm\ge1$), the viscous scale is larger than 
the resistive one, $\lvisc>\lres$. 
As $\Re$ is increased compared to $\Rm$, the viscous scale 
decreases and eventually becomes smaller than the resistive scale. 
At the intermediate values of $\Re$, the resistive scale 
finds itself transiting the spectral bottleneck associated with 
the viscous cutoff until it eventually emerges into the inertial 
range. This transition is illustrated by \figref{fig_spRm}, which shows 
the spectra of the kinetic and magnetic energies for two of the fixed-$\Rm$ 
sequences of runs whose growth rates were shown in \figref{fig_gamma}(a). 
It is obvious that the properties of the velocity field around the 
viscous scale do depend on the type of viscous dissipation employed.
Therefore, the curves $\gamma(\Re,\Rm)$ and $\Rmc(\Re)$ 
cannot be universal around the transitional values of $\Re$ and $\Rm$. 
In particular, since the bottleneck region is narrower in the wave-number 
space for the hyperviscous runs (see \figref{fig_spRm}), 
we expect that so is the transition 
region in the parameter space.\footnote{For the PENCIL-code runs 
in \citeasnoun{SHBCMM_lowPm2}, 
the relatively higher values of $\Rm$ and $\Re$ at which the transition 
seems likely to occur may also be due to the 
generally more dissipative character of a grid code 
compared to the pseudospectral ones. Another possible source of nonuniversality 
might be the difference between the forcing schemes --- if the dynamo is controlled 
by the outer-scale motions, a possibility discussed at the end of 
\secref{sec_faqs}.} 

The degree to which the properties of the magnetic field in 
the Laplacian and hyperviscous runs become similar in the limit 
$\Re\gg\Rm\gg1$ can be judged from \figref{fig_spRe}, where we show 
the magnetic-energy spectra corresponding to the same value of 
$\Re\sim 4000$ and four different values of $\Pm$. For each value 
of $\Pm$, the spectra obtained in a Laplacian and a hyperviscous 
run are given. As $\Pm$ decreases, the bulk of the magnetic energy 
is separated in the wave-number space from the nonuniversal 
viscous cutoff and the magnetic-energy spectra in the Laplacian 
and hyperviscous runs resemble each other more. 

\Figref{fig_cuts_comp} gives a visual illustration 
of the Laplacian vs.\ hyperviscous simulations at low $\Pm$. 
We show snapshots of the velocity and magnetic field in 
two such runs (C4 and HC3) with similar values of $\Rm\sim200$ 
and $\Re\sim4000$. 
While some differences in the velocity structure are visible, 
the magnetic fields look very similar. 

The growth/decay rates for the Laplacian and hyperviscous runs 
with similar $\Rm$ and $\Re$ are also quite close. 
The entire curves $\gamma(\Pm)$ 
can only be meaningfully compared for two of 
the run sequences shown in \figref{fig_gamma}(a): 
$\Rm\sim110$ (series A vs.\ series HA) and 
$\Rm\sim230$ (series B vs.\ series HB). 
Away from the transition region, they appear to agree quite well. 
Further evidence in favour of the equivalence of the Laplacian 
and hyperviscous runs at $\Re\gg\Rm\gg1$ is provided by the 
agreement between saturated energies of the induced magnetic 
fluctuations in such runs (see \secref{sec_highRm}).

\begin{figure}[p]
\begin{tabular}{ccc}
\psfig{file=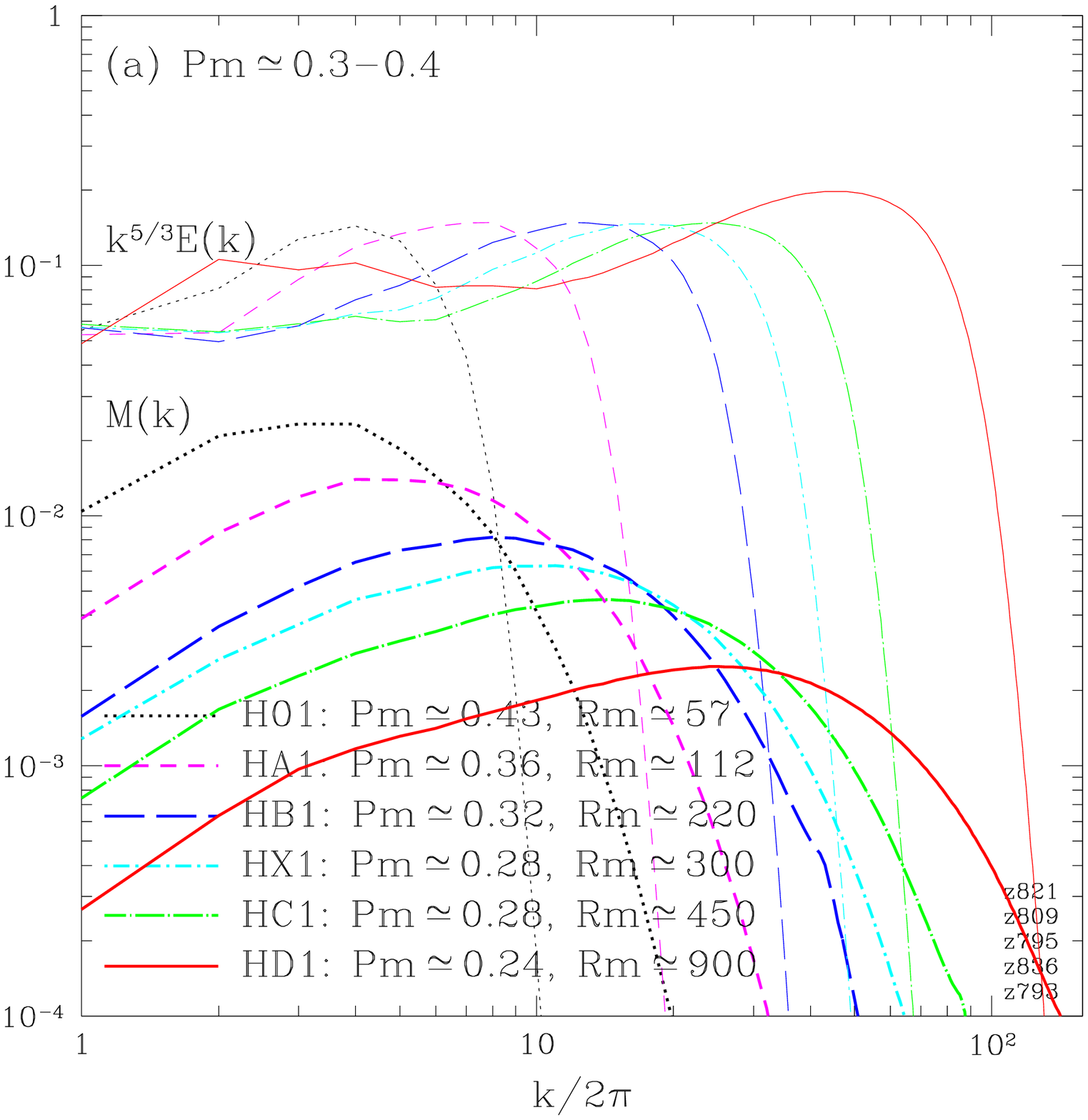,width=7.5cm} && \psfig{file=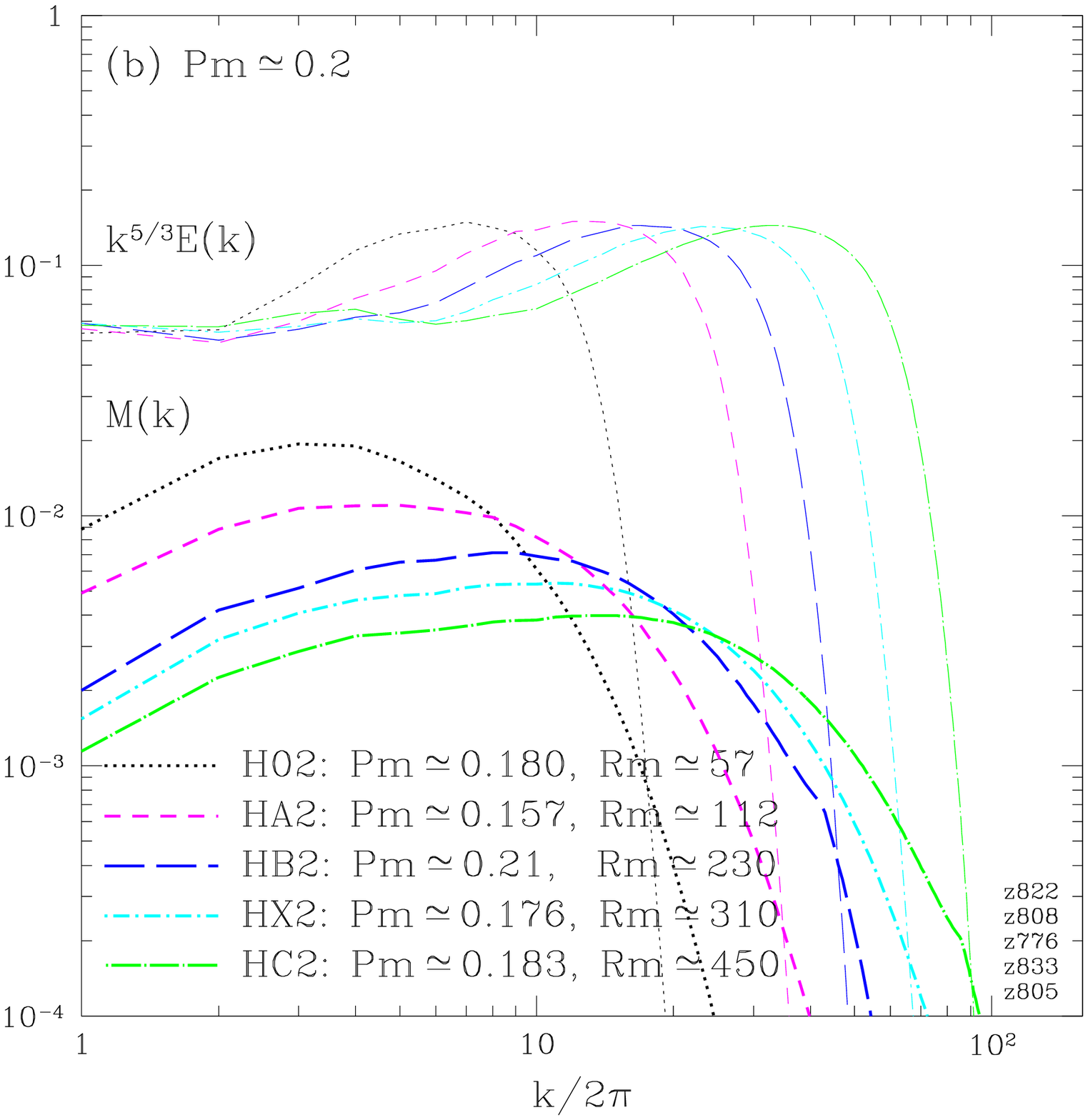,width=7.5cm}\\ 
\psfig{file=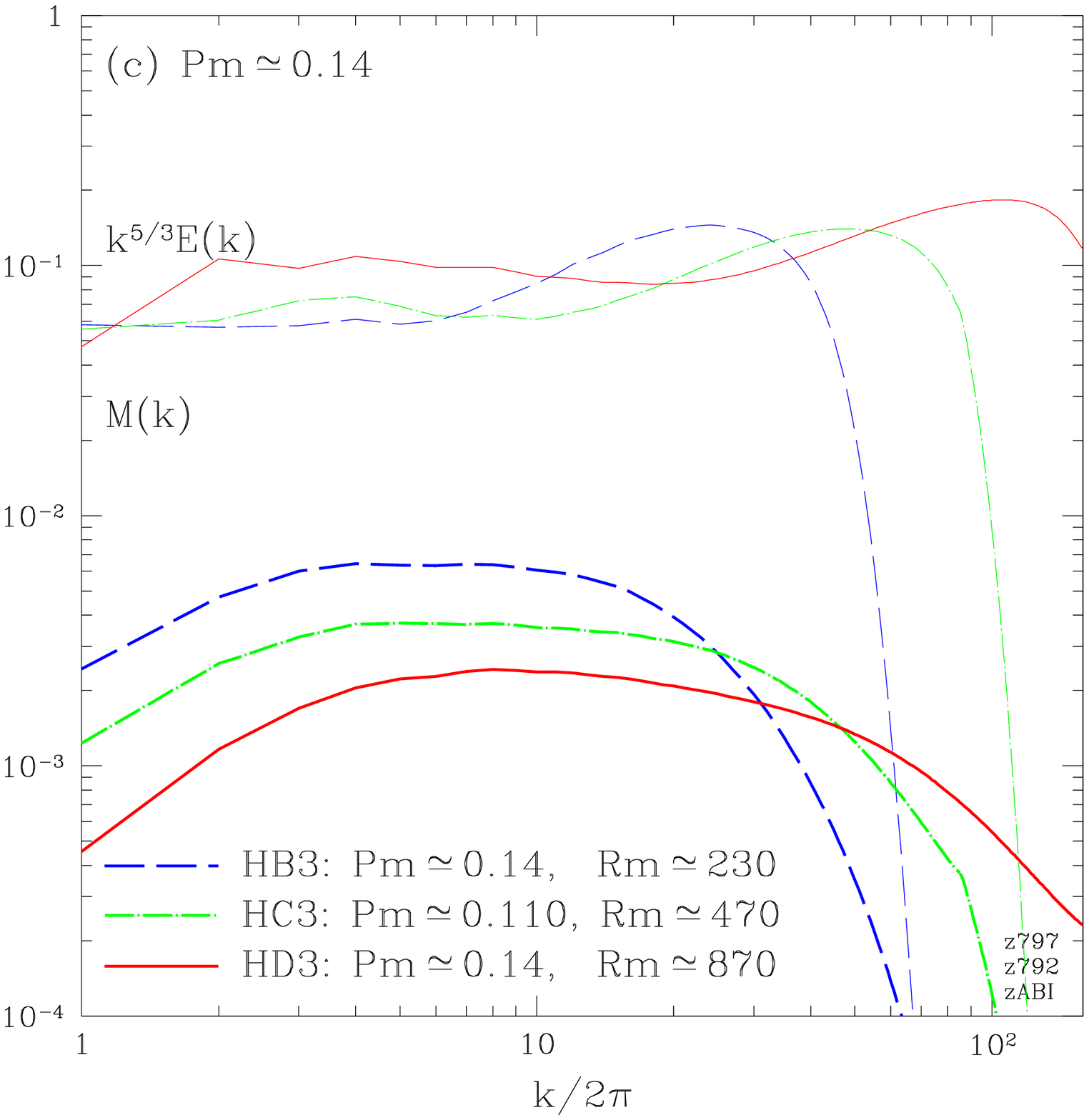,width=7.5cm} && \psfig{file=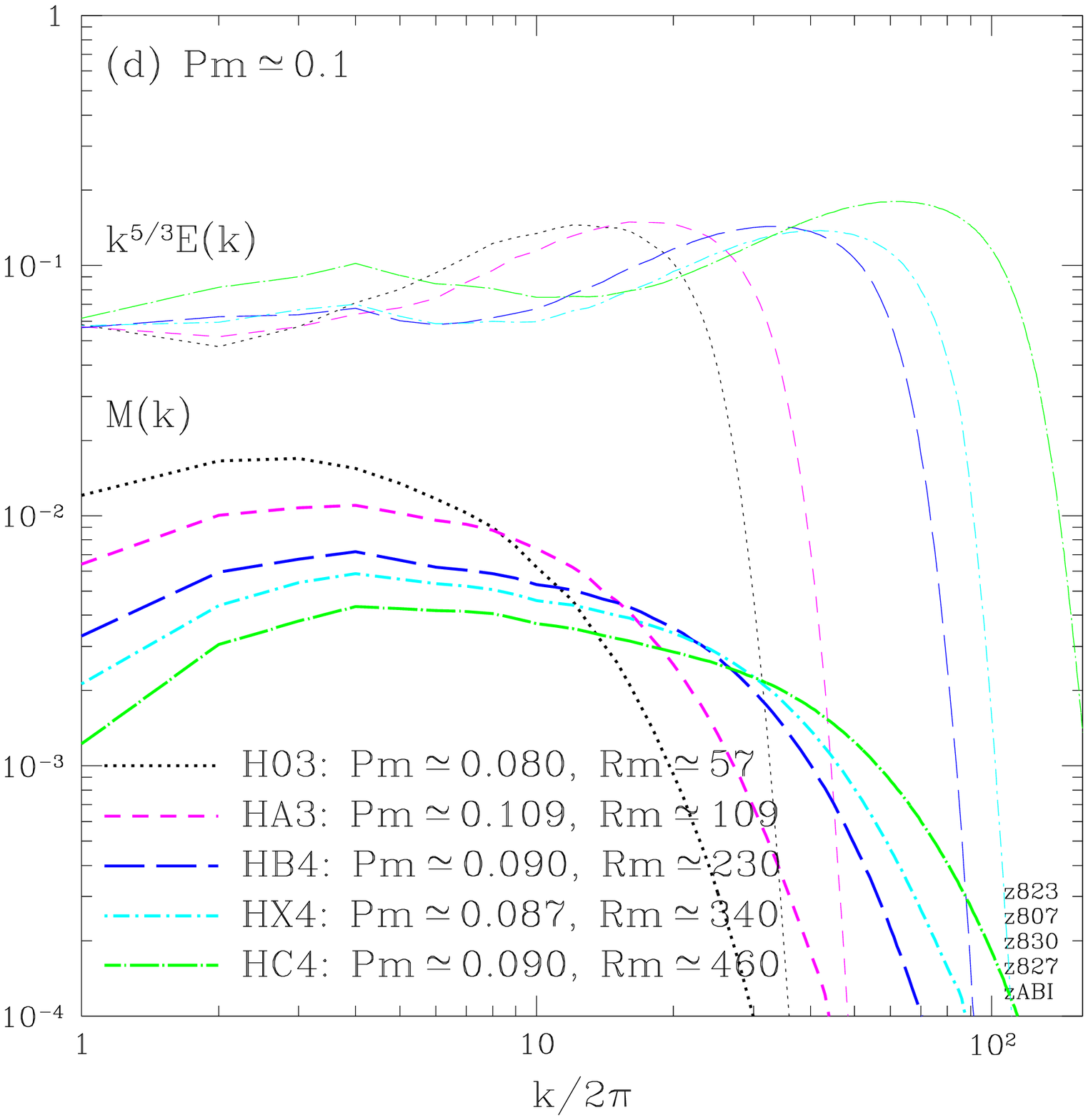,width=7.5cm}\\
\psfig{file=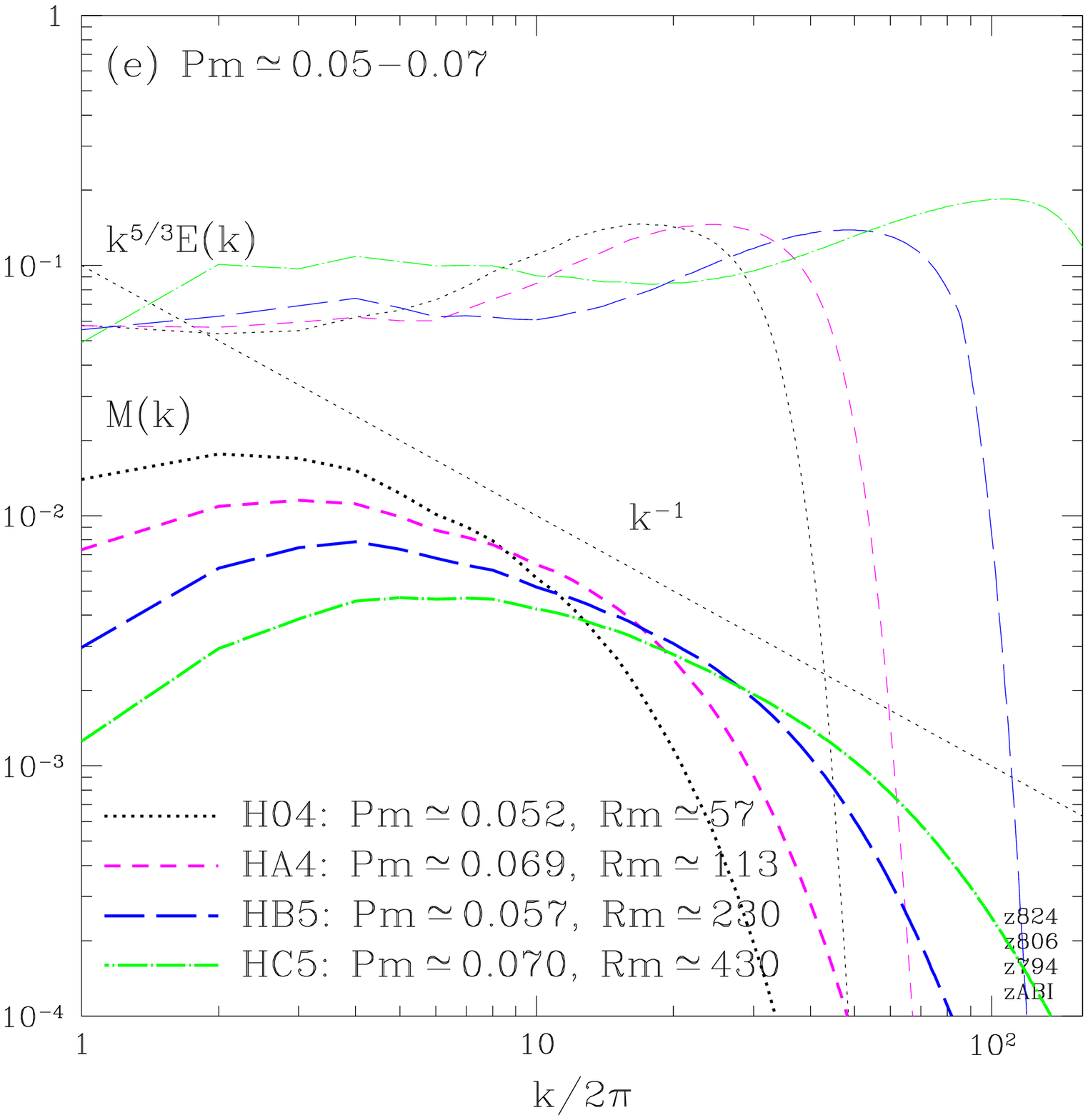,width=7.5cm}   && \psfig{file=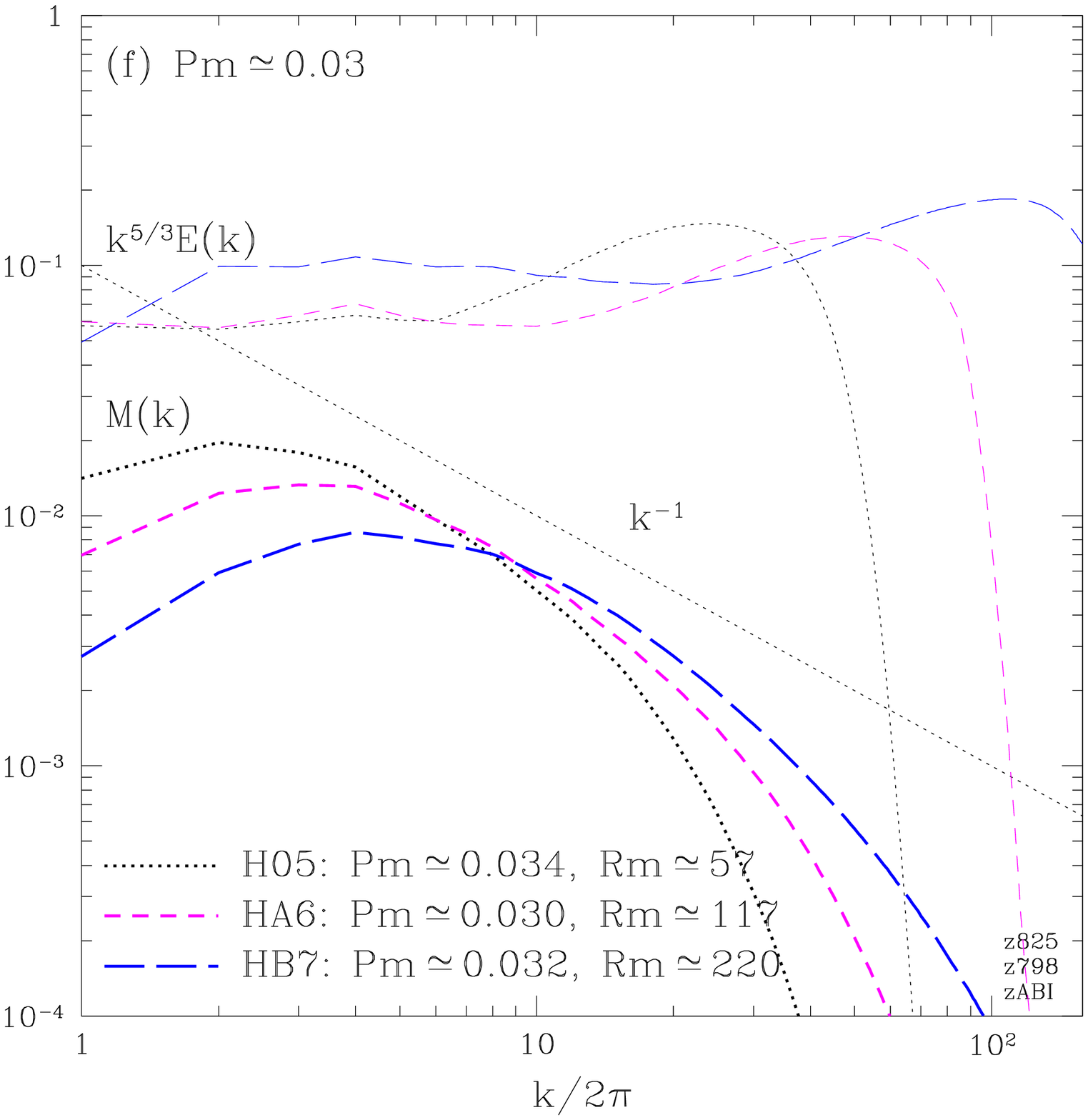,width=7.5cm}
\end{tabular}
\caption{\label{fig_spPm} Spectra of the kinetic and magnetic 
energies (normalized as in \figref{fig_spRm}) 
for the kinematic decaying/growing cases with 
(approximately) fixed $\Pm$: 
(a) $\Pm\simeq0.3-0.4$, (b) $\Pm\simeq0.2$, (c) $\Pm\simeq1.4$, (d) $\Pm\simeq0.1$, 
(e) $\Pm\simeq0.05-0.07$, (f) $\Pm\simeq0.03$. 
The $k^{-1}$ slope is given for reference and discussed in \secref{sec_ind}.} 
\end{figure}

\subsection{Results: Magnetic-Energy Spectra}
\label{sec_spectra}

In examining \figref{fig_spRm}, it is 
hard not to notice that the shape of the magnetic-energy spectrum changes 
as $\Re$ is increased. At $\Pm$ above and just below 
unity, the spectrum has a positive slope and its peak is at the 
resistive scale. This is a typical situation for the fluctuation 
dynamo at $\Pm\ge1$ --- in the limit $\Pm\gg1$, a $k^{+3/2}$ spectrum 
is expected, known as the \citeasnoun{Kazantsev}, or \citeasnoun{KA}, spectrum. 
As the system enters the low-$\Pm$ regime, 
the spectral slope flattens and then becomes negative. 
Since this is a qualitative change 
and since, as far as we know, such spectra have not been seen before, 
it is worth documenting them in more detail. 

The series of plots presented in \figref{fig_spPm} illustrates how 
the magnetic-energy spectrum depends on $\Pm$. In each of these plots, 
we have assembled together the spectra for runs 
with different $\Rm$ and $\Re$ whose ratio was approximately equal 
to a chosen fixed value of $\Pm$ --- approximately because 
for the hyperviscous case, one cannot fix $\Pm$ exactly 
before performing the simulation (see \eqsand{nueff_def}{Re_def}). 
Making allowances for this imprecision, it is still possible 
to conclude tentatively from \figref{fig_spPm} 
that the slope of the magnetic-energy 
spectrum depends on $\Pm$ but not on individually on $\Re$ or $\Rm$. 
As $\Pm$ decreases, the slope turns from positive to negative. 
The data appears more consistent with the peak of 
the spectrum shifted towards the outer scale 
than with it moving with the resistive scale 
($\lres\sim L\Rm^{-3/4}$; see \secref{sec_faqs}), 
but again, this is only a tentative conclusion. 

Unfortunately, here as everywhere else, an attempt to establish 
numerically a solid asymptotic result is frustrated by the resolution 
constraints: in order to determine the spectral slope or the 
position of the spectral peak, we need 
to resolve the limit $\Re\gg\Rm\gg1$ (i.e., both $\Pm\ll1$ and 
$\Rm\gg1$), but we cannot currently achieve sufficiently large values 
of $\Rm$ for $\Pm\ll1$. We shall, therefore, not make any final statements 
here about the asymptotic form of the magnetic spectrum, although 
in \figref{fig_spPm}(e,f) we did provide the reference slope of $k^{-1}$ 
and will discuss it as a theoretical possibility in \secref{sec_highRm}. 

Note finally that the form of the spectrum does not change 
qualitatively between the growing and decaying runs: for example, 
in \figref{fig_spPm}(c), the magnetic energy in run HB3 decays 
while in runs HC3 and HD3 it grows, but the spectral slope appears 
to be the same. 

\subsection{Discussion: Relation to Results for Turbulence with a Mean Flow} 
\label{sec_meanflow}

Since we have claimed above that the low-$\Pm$ fluctuation dynamo had not 
previously been seen in numerical simulations, it is 
important to explain how our results should be compared with 
the $\Rmc(\Re)$ dependences obtained in the recent numerical 
studies by \citeasnoun{Nore_etal}, 
\citeasnoun{Ponty_etal1}, \citeasnoun{Mininni_etal}, \citeasnoun{Ponty_etal2}, 
\citeasnoun{Laval_etal}, \citeasnoun{Mininni_Montgomery},  
\citeasnoun{Mininni_PoP} and \citeasnoun{Mininni}.
In their simulations, turbulence was forced not by a random 
large-scale white noise, but by an imposed body force constant in time
(a Taylor-Green forcing in the first five references, an ABC forcing 
in the other three). This produces a mean flow, 
i.e., a constant (mostly large-scale) spatially inhomogeneous 
velocity field that persists under averaging over times far exceeding 
its own turnover time. There is also 
a fluctuating multiscale velocity field (turbulence), which 
coexists with the mean flow and 
is energetically a few times weaker than it. 
Together, the mean flow plus the turbulence are 
a nonlinear solution of the Navier-Stokes equation \eqref{u_eq} with 
constant forcing. The primary motivation for studying the dynamo 
properties of such a field is its close resemblance to the velocity field 
in liquid-metal dynamo experiments 
(e.g., \citeasnoun{Peffley_etal}, \citeasnoun{Spence_etal}, 
\citeasnoun{Monchaux_etal}; the dynamo properties of 
the flows specific to these experiments have also been studied 
numerically: see, e.g., \citeasnoun{Ravelet_etal}, 
\citeasnoun{Bayliss_etal}).

The mean flows that develop in such systems are usually dynamos 
by themselves --- {\em mean-field dynamos}, to be precise 
(both in the case of helical mean flows like the ABC and in the nonhelical 
case of the Taylor-Green forcing). They give rise to 
growing magnetic fields at scales larger than the 
scale of the flow (or comparable to it, when the scale of the mean flow 
is comparable to the size of the simulation box). 
For $\Pm\ge1$, the threshold for the field amplification 
is very low in these systems: $\Rmc\sim10$, which is a typical situation 
for the mean-field dynamos \citeaffixed{Galanti_etal,Brandenburg}{cf.}. 
The presence of a large amount of the small-scale magnetic energy in 
these simulations should most probably be attributed to the {\em magnetic 
induction}, rather than to the fluctuation dynamo, because the growth 
is happening at values of $\Rm$ that are well below the fluctuation-dynamo 
threshold ($\Rmc\sim60$ for $\Pm\ge1$; see \figref{fig_gamma}(b)). 

As $\Pm$ is decreased, $\Rmc$ increases 
and eventually saturates at some larger value, giving rise 
to a stability curve $\Rmc(\Re)$ that looks similar 
to the stability curve we have obtained in our simulations. 
This similarity (enhanced sometimes by the ambiguity in the 
definitions of the Reynolds numbers) ought not to lead to any 
confusion between the two curves. In order to illustrate the 
difference between them, we have plotted in \figref{fig_gamma}(b) 
the stability curve reported by \citeasnoun{Ponty_etal2} for their 
simulations with a Taylor-Green forcing. The data is taken from their table 1 
and calibrated according to our definitions of $\Re$ and $\Rm$: 
they define $\Re=\sqrt{\usq}\, L_{\rm dyn}/\nu$, where $L_{\rm dyn}$ is 
the dynamical integral scale computed from the kinetic-energy 
spectrum and given in their table 1, while we define 
$\Re$ according to \eqref{Re_def}. The box wave number 
$k_0=1$ in their calculations. 
We see that the two curves are very different: the 
dynamo threshold for the simulations with a mean flow 
is much lower than for our homogeneous simulations. 
The difference is not merely quantitative: 
the ordered large-scale structure of the growing magnetic 
field in the $\Pm\sim1$ runs of \citeasnoun{Ponty_etal1} 
(the lower part of their stability curve) confirms that 
it is a mean-field dynamo. 

The increase of $\Rmc$ with increasing 
$\Re$ in these simulations has been attributed to the 
interference by the turbulence with the dynamo properties 
of the mean flow --- a manifestation of a tendency 
for higher dynamo thresholds in the presence of 
of a large-scale noise \cite{Petrelis_Fauve,Laval_etal}. 
It would be interesting to check whether the turbulence in 
these simulations might itself act as a dynamo if the 
mean flow is ``manually'' removed from the induction 
equation \eqref{B_eq}. Comparison of the two stability 
curves in \figref{fig_gamma}(b) suggests that in order for this 
to be the case, $\Rm$ must be increased very substantially 
--- above the threshold found by us. 

Let us conclude this discussion on a speculative note. 
While simulations with a mean flow undoubtedly exhibit 
a mean-field dynamo at $\Pm\ge1$, it is not clear whether 
that is also the case when $\Pm\ll1$. 
It may or may not be a coincidence that the low-$\Pm$ threshold for the 
simulations with a mean flow appears to be very close to the 
large-$\Pm$ fluctuation-dynamo threshold. 
Since the mean flow most probably has chaotic Lagrangian trajectories 
(although this has not been explicitly checked), it should be 
a fluctuation dynamo --- it belongs to the same class as the large-$\Pm$ 
dynamos because the flow is spatially smooth. 
Could the turbulence, while suppressing 
the mean-field dynamo in the low-$\Pm$ regime,  
somehow fail to interfere with the fluctuation dynamo 
of the mean flow? There is not enough evidence or physical 
insight available to us to judge the merits of this possibility --- 
or of another, equally speculative one that we float at the end of 
\secref{sec_faqs}. Two aspects of the published numerical results 
that seem to support it are the evidence of direct nonlocal 
transfer of energy from the outer scale (the mean flow) to the 
small-scale magnetic field \cite{Mininni_Alexakis_Pouquet} and the 
fact that the magnetic-energy spectra 
reported by \citeasnoun{Ponty_etal1} do not exhibit the 
tendency towards a negative slope discussed in \secref{sec_spectra}, 
but rather bear a strong resemblance to the typical $k^{+3/2}$ 
spectra of the fluctuation dynamo at $\Pm\ge1$ 
\citeaffixed{SCTMM_stokes,Haugen_Brandenburg_Dobler}{see, e.g.,}. 

\subsection{Discussion: Relation to Theory and Outstanding Questions}
\label{sec_faqs}

Given the numerical certainty that the low-$\Pm$ fluctuation dynamo 
exists, there remain a number of theoretical uncertainties 
about the nature of this dynamo. The key question is whether or not 
it is the inertial-range motions that amplify the magnetic energy. 

Suppose $\Re\gg\Rm\gg1$. 
Let us examine what can be achieved by assuming that the transfer 
of the kinetic into magnetic energy is {\em local} in scale space.\footnote{Whether 
it really is local should be the subject of a thorough future investigation, possibly 
along the lines of the recent study by \citeasnoun{Mininni_Alexakis_Pouquet} 
of a mean-flow-driven dynamo. Note that the shell-model simulations 
recently carried out by \citeasnoun{Stepanov_Plunian}, which
enforce the locality of interactions, return a picture that 
seems to be broadly in agreement with the inertial-range dynamo 
scenario discussed below. On the other hand, EDQNM closure simulations 
do not find any increase in $\Rmc$ for $\Pm\ll1$ compared with $\Pm\ge1$ 
\cite{Leorat_Pouquet_Frisch}.} 
For Kolmogorov turbulence, the characteristic velocity 
fluctuation at scale $l$ is $\du_l\sim (\eps l)^{1/3}$, 
where $\eps$ is the total power 
injected into the turbulence (the turbulent energy flux).
The characteristic rate of stretching of the 
magnetic field by the velocity field at scale $l$ is then
$\du_l/l\sim \eps^{1/3}\l^{-2/3}$. 
The characteristic rate of turbulent diffusion at scale $l$ 
is of the same order. Comparing the stretching rate with the rate 
of the Ohmic diffusion of the magnetic field, $\eta/l^2$, one finds 
the resistive scale, i.e., the scale at which 
the stretching rate is maximal and below which 
it is overcome by diffusion \cite{Moffatt}: 
\bea
\label{lres_def}
\lres\sim \lt({\eta^3\over\eps}\rt)^{1/4} \sim {\lf\over\Rm^{3/4}}
\eea
(this resistive scale does, indeed, lie inside the 
inertial range, $\lf\gg\lres\gg\lvisc$, where 
the viscous scale $\lvisc$ is estimated using 
\eqref{lres_def} with $\eta$ replaced by $\nu$). 
Thus, if the local interaction of the inertial-range motions 
with the magnetic field is capable of amplifying the field in a 
sustained way, the growth rate of the magnetic energy should scale with $\Rm$ as 
\bea
\label{gamma_Rm}
\gamma \sim {\du_{\lres}\over\lres} \sim {\eps^{1/3}\over\lres^{2/3}} 
\sim \lt({\eps\over\eta}\rt)^{1/2} \sim {\urms\over\lf}\,\Rm^{1/2} 
\eea 
in the limit $\Re\to\infty$. 
This dynamo, if it exists, is purely a property of the inertial 
range and is independent of any system-dependent 
outer-scale circumstances such as, e.g., the presence of a mean flow. 
Given a large enough $\Rm$, the growth rate \eqref{gamma_Rm} will 
always be larger than a mean-field or any other kind of dynamo 
associated with the outer-scale motions, because the latter 
cannot amplify the field faster than at the rate 
\bea
\label{gamma_mf}
\gamma \sim {\umf\over\lf} \sim {\urms\over\lf},
\eea
where $\umf$ is the characteristic velocity at the outer scale. 

While our numerical results allowed us to make what we consider 
a compelling case for the existence of a positive asymptotic 
value of the growth rate (\secref{sec_Rmc}), 
we cannot at this stage check whether it scales with $\Rm$ 
according to \eqref{gamma_Rm} or reaches an $\Rm$-independent 
limit as in \eqref{gamma_mf}. Neither of these possibilities 
can be ruled out {\em a priori}.

Should the scaling \eqref{gamma_Rm} 
be confirmed, the case for an inertial-range dynamo would 
be complete. This case is strengthened somewhat 
by the theoretical predictions based on the only available model 
of the turbulent dynamo that is solvable exactly --- the \citeasnoun{Kazantsev} 
model. This model considers a Gaussian random velocity field that 
is a white noise in time. The salient property of the 
inertial-range velocities is their spatial roughness: 
$\du_l\sim l^{1/3}$ for $L\gg l\gg\lvisc$ 
compared with a smooth velocity $\du_l\sim l$ for $l\ll\lvisc$. 
This is mimicked by prescribing a spatially rough power-law 
correlation function for the Kazantsev model field. 
For a certain range of exponents of this power law, it is then 
possible to show analytically that the Kazantsev field 
is a dynamo \cite{Rogachevskii_Kleeorin,Vincenzi,Boldyrev_Cattaneo,Celani_Mazzino_Vincenzi,Arponen_Horvai}. 
The difficulty lies in establishing a quantitative connection between 
this result and the real turbulence, in which the decorrelation time 
of the inertial-range motions is certainly not small but 
comparable to their turnover time and scale-dependent $\sim l^{2/3}$. 
It is not known whether setting it to zero changes the dynamo 
properties of the velocity field enough to render the white-noise 
model irrelevant. If the white-noise velocity is on some level acceptable, 
it is not known what choice of its roughness exponent 
(which determines whether it is a dynamo!) makes it a good 
model of the inertial-range turbulent velocity field. 
The authors cited above used a plausible argument 
\citeaffixed{Vainshtein_lowPm}{first suggested by}, 
which led them to predict dynamo action. However, this 
prediction is purely a {\em quantitative} 
mathematical outcome of analyzing a synthetic velocity field 
that is at best a passable {\em qualitative} representation 
of real turbulence. In the absence of a physical model of the 
inertial-range dynamo, the validity of this prediction 
remains in doubt. 

It is natural to ask whether any of the quantitative 
predictions based on the Kazantsev model are bourne out by our 
numerical results. One such prediction is the scaling \eqref{gamma_Rm} 
of the growth rate, which cannot as yet be verified numerically. 
Another, due to \citeasnoun{Rogachevskii_Kleeorin} and to 
\citeasnoun{Boldyrev_Cattaneo}, is the expectation that the asymptotic 
dynamo threshold $\Rmcinf$ for $\Pm\ll1$ should be approximately 7 times 
higher than a similar threshold $\Rmc\sim60$ for the $\Pm\gg1$ case 
\citeaffixed{Novikov_etal}{see \figref{fig_gamma}(b); in the Kazantsev model, 
the latter threshold is computed by using a velocity field with a smooth spatial 
correlator, see}. This would imply $\Rmcinf\sim400$, which is an overestimate 
by at least a factor of 2 (see \secref{sec_Rmc}) --- certainly not a damning 
contradiction, but somewhat short of a confirmation of the theory. 
Finally, \citeasnoun{Boldyrev_Cattaneo} predict a 
(stretched) exponential fall off of the magnetic-field 
correlation function at $l>\lres$, so the magnetic energy is 
concentrated sharply at the resistive scale. This appears to 
be at odds with the trend for the magnetic-energy spectrum to develop 
a negative slope above the resistive scale, reported in \secref{sec_spectra}
(for example, a $k^{-1}$ spectrum would imply that $\dB_l\sim\const$, 
i.e., the correlation function is flat rather than falling off exponentially). 
We reiterate, however, that at resolutions currently available to us, 
we are unable to claim definitively that the spectral peak is not, in fact, 
at the resistive scale. 

Is there an alternative to the inertial-range dynamo? 
It might be worth asking whether the randomly 
forced outer-scale motions could act as a dynamo 
despite (or in concert with) the turbulence in 
the inertial range. Indeed, how essential is the physical difference 
between the outer-scale motion, whose decorrelation time 
($\sim\lf/\umf$) is long compared with the inertial-range 
motions, and a mean flow, whose correlation time in infinite? 
Can both the mean-flow-driven dynamo found by \citeasnoun{Ponty_etal1} 
and the fluctuation dynamo reported here by us be manifestations 
of some universal basic mechanism --- for example, of the field amplification 
by a combined action of a persistent 
(slowly changing) large-scale (outer-scale) shear 
and small-scale (inertial-range) turbulent fluctuations
\citeaffixed{RK03}{a nonhelical mean-field dynamo of this type has been 
proposed theoretically by}? 

An unambiguous signature of this or any other 
type of outer-scale-driven dynamo would be the convergence 
of the growth rate to an $\Rm$-independent limit \eqref{gamma_mf}. 
One way to investigate numerically whether there is 
a smooth connection between the mean-flow dynamo and 
the randomly forced one would be to construct stability 
curves $\Rmc(\Re)$ for 
a series of numerical experiments with an inhomogeneous 
body force \citeaffixed{Ponty_etal1}{similar to}, which however, 
is artificially decorrelated with a prescribed correlation 
time $\tcorr$. The limit $\tcorr=\infty$ corresponds 
to the turbulence with a mean flow. As $\tcorr$ is reduced, 
the mean flow should develop a slow time dependence and 
at $\tcorr\sim\lf/\umf$, the situation would become equivalent 
to the randomly forced case discussed here. 

\begin{table}[t]
\caption{\label{tab_ind} Index of runs --- Part III (runs with a mean field)}
\lineup
{\footnotesize
\begin{tabular}{@{}lrcrllrrlll}
\br
Run    &Res.   & $\Bo$     &     $\eta\qquad$& \Rm & $\Pm$ & $\Re$&$\Rel$&$\urms$& $\dBrms$& Fig.\\
\mr
\multicolumn{11}{l}{$\nu_2=5\times10^{-4}$}\\
\mr
C1-sat &$256^3$& 0         & $5\times10^{-4}$& 390 & 1.0   &  390 &149 & 1.22  & 0.49    &\ref{fig_E}(b) \z{583}\\
B2-sat &$256^3$& 0         &        $10^{-3}$& 200 & 0.5   &  410 &149 & 1.28  & 0.42    &\ref{fig_E}(b) \z{706}\\
m0     &$128^3$& $1$       & $2\times10^{-3}$& 103 & 0.25  &  410 &169 & 1.29  & 0.74    &\ref{fig_E}(a) \z{697}\\
m1     &$128^3$& $10^{-1}$ & $2\times10^{-3}$&  98 & 0.25  &  390 &161 & 1.24  & 0.54    &\ref{fig_E}(a) \z{696}\\
m2     &$128^3$& $10^{-2}$ & $2\times10^{-3}$& 109 & 0.25  &  440 &118 & 1.37  & 0.136   &\ref{fig_E}(a) \z{669}\\
m3/M1.1&$128^3$& $10^{-3}$ & $2\times10^{-3}$& 109 & 0.25  &  440 &111 & 1.37  & 0.021   &\ref{fig_E}(a,b) \z{667}\\
m4     &$128^3$& $10^{-4}$ & $2\times10^{-3}$& 112 & 0.25  &  450 &114 & 1.41  & 0.0025  &\ref{fig_E}(a) \z{668}\\
m5     &$128^3$& $10^{-5}$ & $2\times10^{-3}$& 112 & 0.25  &  450 &114 & 1.41  & 0.000165&\ref{fig_E}(a) \z{698}\\
M1.2   &$128^3$& $10^{-3}$ & $4\times10^{-3}$&  57 & 0.125 &  460 &113 & 1.43  & 0.0066  &\ref{fig_E}(b) \z{702}\\
M1.3   &$128^3$& $10^{-3}$ &        $10^{-2}$&  23 & 0.05  &  450 &115 & 1.42  & 0.0030  &\ref{fig_E}(b) \z{703}\\
M1.4   &$128^3$& $10^{-3}$ & $2\times10^{-2}$&11.6 & 0.025 &  470 &117 & 1.46  & 0.0020  &\ref{fig_E}(b) \z{704}\\
M1.5   &$128^3$& $10^{-3}$ & $4\times10^{-2}$& 5.5 & 0.0125&  440 &110 & 1.39  & 0.00132 &\ref{fig_E}(b) \z{705}\\
M1.6   &$128^3$& $10^{-3}$ &        $10^{-1}$& 2.2 & 0.005 &  440 &115 & 1.39  & 0.00074 &\ref{fig_E}(b) \z{768}\\
M1.7   &$128^3$& $10^{-3}$ & $2\times10^{-1}$& 1.08& 0.0025&  430 &113 & 1.36  & 0.00037 &\ref{fig_E}(b) \z{769}\\
M1.8   &$128^3$& $10^{-3}$ & $4\times10^{-1}$& 0.54&0.00125&  430 &112 & 1.35  & 0.00020 &\ref{fig_E}(b) \z{770}\\
M1.9   &$128^3$& $10^{-3}$ &                1& 0.22& 0.0005&  440 &115 & 1.39  &0.000069 &\ref{fig_E}(b) \z{771}\\
\mr
\multicolumn{11}{l}{$\nu_2=2.5\times10^{-4}$}\\
\mr
D1-sat$^*$&$512^3$& 0     &$2.5\times10^{-4}$& 710 & 1.0   &  710 &185 & 1.11  & 0.53    &\ref{fig_E}(b) \zABI{$(14.4,18.4)$}\\
C2-sat$^*$&$256^3$& 0      & $5\times10^{-4}$& 380 & 0.5   &  760 &210 & 1.20  & 0.45    &\ref{fig_E}(b) \zABI{$(30,50)$}\\
B3-sat &$256^3$& 0         &        $10^{-3}$& 210 & 0.25  &  840 &230 & 1.30  & 0.35    &\ref{fig_E}(b) \z{588}\\
M2.1   &$256^3$& $10^{-3}$ & $2\times10^{-3}$& 105 & 0.125 &  840 &157 & 1.32  & 0.0088  &\ref{fig_E}(b) \z{701}\\
M2.2   &$256^3$& $10^{-3}$ & $4\times10^{-3}$&  53 & 0.0625&  840 &155 & 1.33  & 0.0053  &\ref{fig_E}(b) \z{693}\\
M2.3   &$256^3$& $10^{-3}$ &        $10^{-2}$&  21 & 0.025 &  830 &156 & 1.31  & 0.0030  &\ref{fig_E}(b) \z{695}\\
\mr
\multicolumn{11}{l}{$\nu_2=1.25\times10^{-4}$}\\
\mr
C3-sat$^*$&$256^3$& 0      & $5\times10^{-4}$& 380 & 0.25  & 1530 &350 & 1.20  & 0.47    &\ref{fig_E}(b) \zABI{$(30.50)$}\\
M3.1   &$256^3$& $10^{-3}$ &        $10^{-3}$& 210 & 0.125 & 1690 &230 & 1.33  & 0.023   &\ref{fig_E}(b) \z{772}\\
M3.2   &$256^3$& $10^{-3}$ & $2\times10^{-3}$& 111 & 0.0625& 1770 &230 & 1.39  & 0.0132  &\ref{fig_E}(b) \z{767}\\
\mr
\multicolumn{11}{l}{$\nu_8=10^{-20}$}\\
\mr
HM1    &$256^3$& $10^{-3}$ &        $10^{-3}$& 230 & 0.062 & 3700 &340 & 1.45  & 0.027   &\ref{fig_E}(b), \ref{fig_HM} \z{799}\\
HM2    &$256^3$& $10^{-3}$ & $2\times10^{-3}$& 114 & 0.031 & 3600 &330 & 1.43  & 0.0109  &\ref{fig_E}(b), \ref{fig_HM} \z{802}\\
HM3    &$256^3$& $10^{-3}$ & $4\times10^{-3}$&  56 & 0.0160& 3500 &320 & 1.41  & 0.0071  &\ref{fig_E}(b), \ref{fig_HM} \z{803}\\
HM4    &$256^3$& $10^{-3}$ &        $10^{-2}$&  22 & 0.0065& 3400 &310 & 1.39  & 0.0034  &\ref{fig_E}(b), \ref{fig_HM} \z{800}\\
HM7    &$256^3$& $10^{-3}$ &        $10^{-1}$& 2.3 &0.00063& 3600 &330 & 1.43  & 0.00067 &\ref{fig_E}(b), \ref{fig_HM} \z{801}\\
\br
\end{tabular}
}
\end{table}

\section{Turbulent Induction}
\label{sec_ind}

The turbulent magnetic induction is the tangling of 
a uniform (or large-scale) mean magnetic field by turbulence,  
which produces magnetic energy at small scales. 
The mean field may be due to some external 
imposition or to a mean-field dynamo. 
Broadly speaking, whatever is the mechanism that 
generates and/or maintains a magnetic field, 
the turbulent induction is a nonlocal 
energy transfer process whereby this field couples 
to motions at smaller scales to give rise to magnetic 
fluctuations at those scales. In any real system, 
it is only a part of a bigger picture of how the magnetic 
field is generated and shaped. Given a multiscale observed or 
simulated magnetic field, 
one does not generally have enough information (or understanding), to 
tell whether it has originated from the fluctuation dynamo, from 
the mean-field dynamo plus the turbulent induction or from some combination of the two. 
However, in the computer, the turbulent-induction effect 
can be isolated by measuring the response to an
imposed uniform field in the subcritical cases, in which the magnetic field 
would otherwise decay \citeaffixed{Odier_Pinton_Fauve,Bourgoin_etal,Spence_etal}{this
approach has been popular in liquid-metal experiments; see, e.g.,}. 
\Tabref{tab_ind} details a number of such 
runs, done using the same numerical set up as that detailed 
in \secref{sec_setup}. 
We shall see that examining their properties is both 
instructive in itself and may be revealing about the nature 
of the fluctuation dynamo.

\begin{figure}[t]
\begin{center}
\begin{tabular}{ccc}
\psfig{file=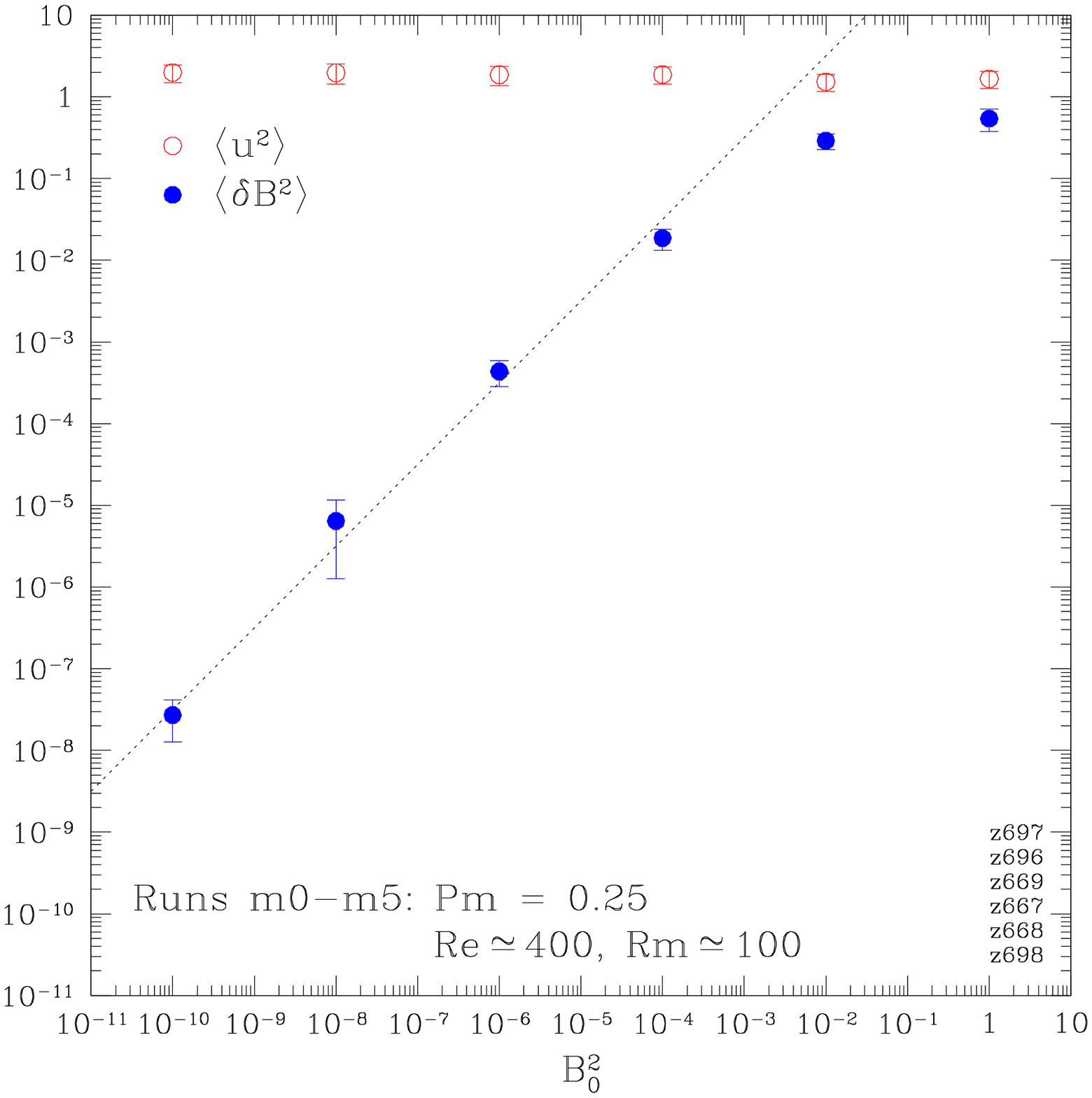,width=7.5cm}&&\psfig{file=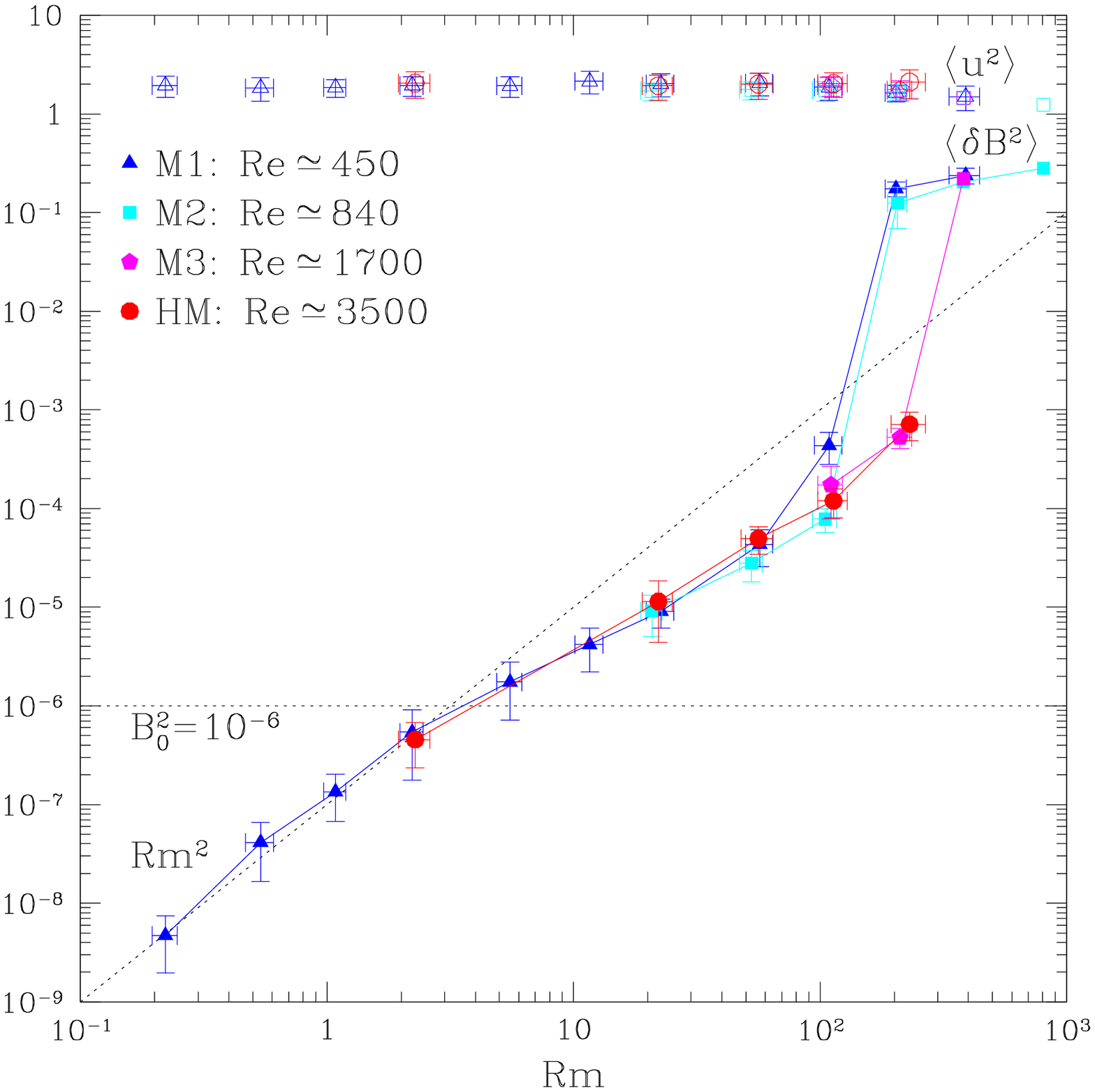,width=7.5cm}\\
(a) && (b)
\end{tabular}
\end{center}
\caption{\label{fig_E} (a) The mean square induced field $\dBsq$ 
vs.~the mean field squared $\Bo^2$ in the subcritical regime 
(no dynamo). These runs correspond to the decaying run A3 
(\tabref{tab_dynamo1}) with an imposed mean field. 
Error bars show the mean square deviation from the average 
value in the set of snapshots over which the time average was done. 
The dotted line shows the slope corresponding to a linear dependence.
(b) The mean square induced field $\dBsq$ vs.~$\Rm$ for runs with $\Bo=10^{-3}$ 
in the subcritical regime, $\Rm<\Rmc$. 
For $\Rm>\Rmc$, the saturated energies for $\Bo=0$ runs are shown 
(the saturated states of runs B2, B3, C1, C2, C3 are denoted B2-sat, etc.\ in 
\tabref{tab_ind}). The reference slope corresponding to $\Rm^2$ is shown
for comparison with the quasistatic theory (\secref{sec_lowRm}).} 
\end{figure}

Mathematically, if 
the magnetic field is represented as a sum of a uniform mean 
field and a fluctuating part, $\vB=\vB_0 + \dvB$, 
the fluctuating field satisfies
\bea
\label{mag_ind}
{\dd\dvB\over\dd t} + \vu\cdot\vdel\dvB = \dvB\cdot\vdel\vu 
+ \eta\nabla^2\dvB + \vB_0\cdot\vdel\vu.
\eea
This is simply the induction equation \eqref{B_eq} with 
a source term, which, in the absence of dynamo action, 
will give rise to a saturated level of the magnetic-fluctuation 
energy. \Eqref{mag_ind} is linear in $\dvB$. 
Therefore, for dynamically weak mean fields, 
the saturated magnetic energy should be proportional 
to the mean-field energy: 
\bea
\dBsq = f(\Rm)\Bo^2.
\label{dB_lin}
\eea 
The coefficient of proportionality $f(\Rm)$ in this relation 
is expected to be an increasing function of $\Rm$. At large $\Rm$, this 
coefficient can be large, $f(\Rm)\gg1$ and the relation \eqref{dB_lin} 
will hold only as long as not only the mean field but also 
the fluctuating field is dynamically 
weak, $\dBsq\ll\usq$. If it is not, the back reaction will take over 
as the controlling agent in the saturation mechanism, resulting 
in a dynamical state with $\dBsq\sim\usq$. 
This simple picture is, unsurprisingly, bourne out by the numerical experiment: 
\figref{fig_E}(a) illustrates this point. 

\subsection{Turbulent Induction at Low $\Rm$} 
\label{sec_lowRm}

It is rather straightforward to make a theoretical prediction about the 
form of $f(\Rm)$ in the limit of small $\Rm$. The diffusion term in 
\eqref{mag_ind} dominates the nonlinear terms and the saturated state 
is governed by the ``quasistatic'' balance
\bea
\label{quasist}
\eta\nabla^2\dvB = - \vB_0\cdot\vdel\vu,
\eea
which immediately implies $\dBsq\sim\Rm^2\Bo^2$. 

In the wave-number space, the quasistatic approximation \eqref{quasist} 
returns an explicit form of the angle-integrated, one-dimensional 
magnetic-energy spectrum $M(k)$ in terms 
of the kinetic-energy spectrum~$E(k)$: 
\bea
\eta k^2\dvB_\vk = \rmi(\vk\cdot\vB_0)\vu_\vk
\quad\Rightarrow\quad
M(k) = {\Bo^2\over 3\eta^2}\,{E(k)\over k^2}.
\label{sp_quasist}
\eea
If the kinetic energy spectrum is Kolmogorov, one 
obtains $M(k)\sim \Bo^2\eta^{-2}\eps^{2/3}k^{-11/3}$  
\cite{Golitsyn}. Note that we limit ourselves to the case of 
weak magnetic field 
\citeaffixed[and references therein]{Zikanov_Thess,Knaepen_Kassinos_Carati}{for 
some numerical experiments with a dynamically strong field, see, e.g.,}. 

\begin{figure}[t]
\centerline{\psfig{file=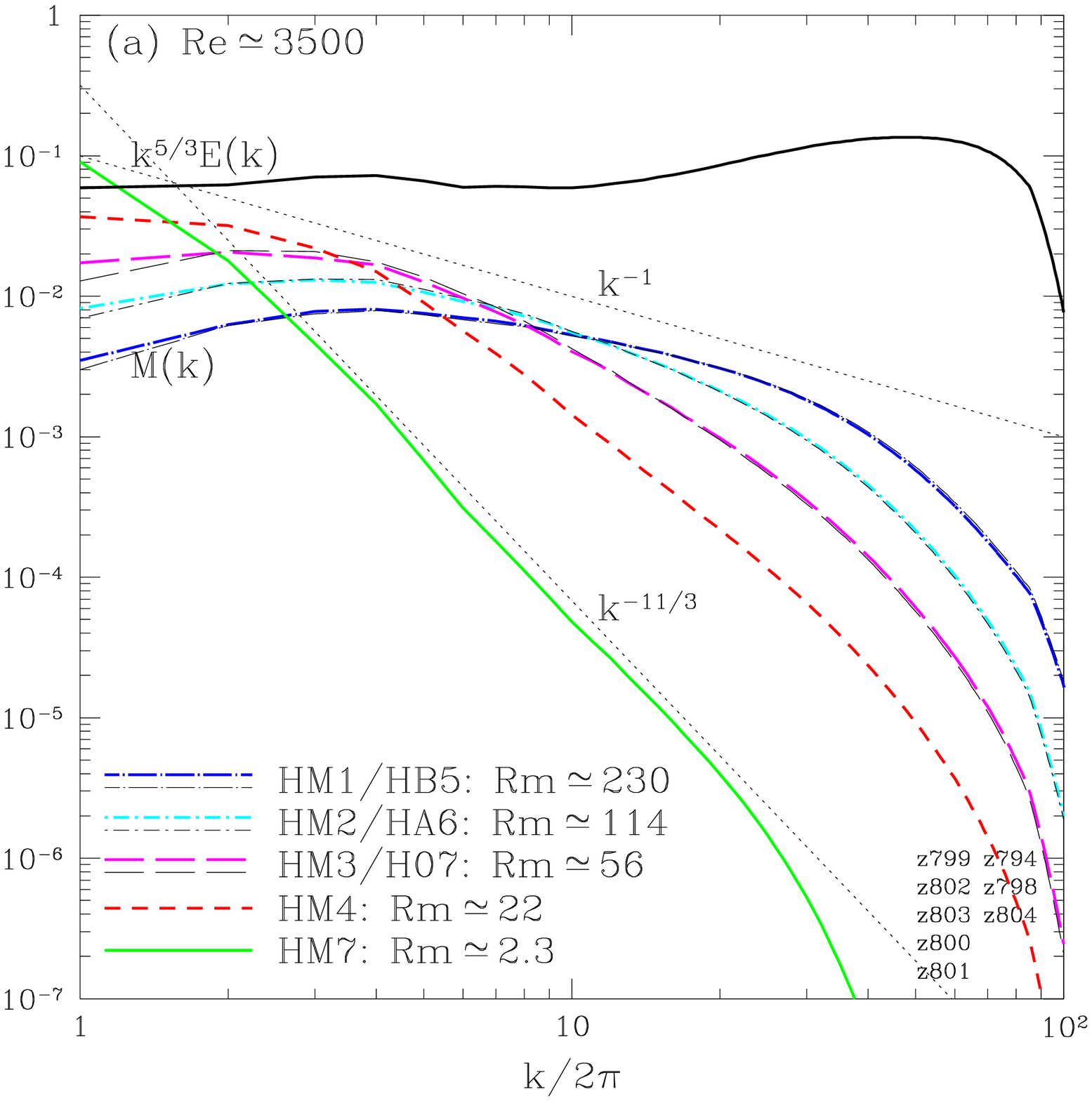,width=7.5cm}\psfig{file=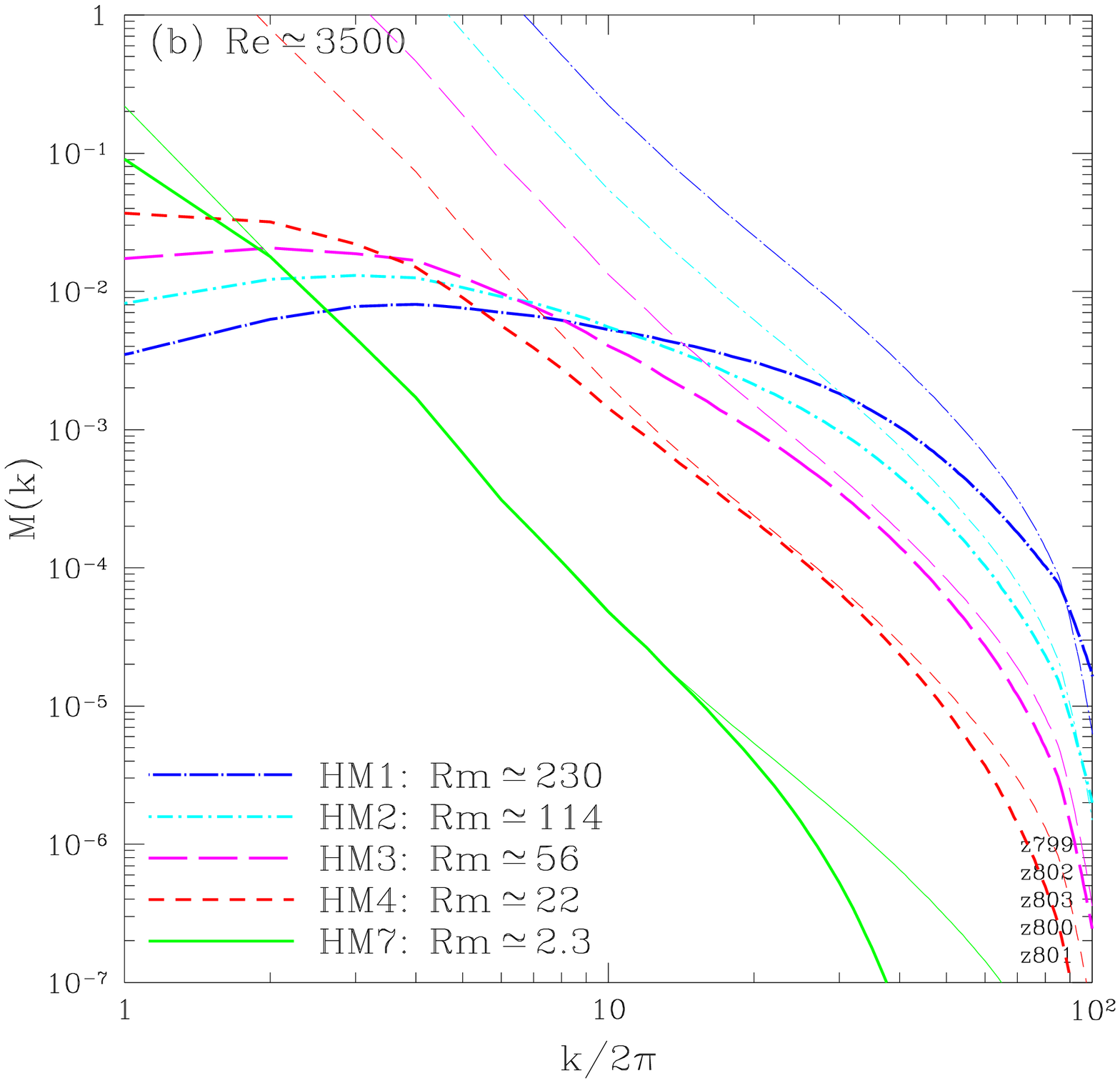,width=7.5cm}}
\caption{\label{fig_HM} (a) The magnetic-energy spectra, 
normalized by $\dBsq/2$ and time-averaged, for the 
runs with the 8th-order hyperviscosity (effective $\Re\sim3500$) and 
$\Bo=10^{-3}$ (series HM). For comparison, normalized magnetic-energy spectra 
for the analogous decaying runs with $\Bo=0$ are shown. The decaying spectra 
(runs HB5, HA6, H07) coincide with the steady-state 
spectra of induced fluctuations (runs HM1, HM2, HM3) almost exactly. 
The slopes corresponding to the \citeasnoun{Golitsyn} 
$k^{-11/3}$ and \citeasnoun{Ruzmaikin_Shukurov} $k^{-1}$ spectra are 
given for reference. The kinetic-energy spectrum, normalized by $\usq/2$, 
time-avearaged and compensated by $k^{5/3}$, 
is also plotted (it is the same for all runs). (b) The same spectra 
as in (a) for the HM series (bold lines) compared with the linear 
quasistatic-theory prediction \eqref{sp_quasist} (thin lines). 
For all runs except HM7, $\Bo$ in \eqref{sp_quasist} has been replaced 
with $\Bsq^{1/2}$.} 
\end{figure}

The predictions of the quasistatic theory are rigourous 
and, of course, confirmed by the 
numerical simulations. \Figref{fig_E}(b) shows that the 
$\Rm^2$ scaling of the induced magnetic energy holds, 
under our definition of $\Rm$, up to $\Rm\sim2$, which is also 
approximately the point at which $\dBsq$ becomes larger than 
$\Bo^2$. The \citeasnoun{Golitsyn} spectrum is also there 
--- as shown in \figref{fig_HM}(a) for the run HM7. 
While this spectrum is solidly established in the 
laboratory \citeaffixed{Odier_Pinton_Fauve,Bourgoin_etal}{e.g.,} 
and was successfully simulated by \citeasnoun{Ponty_Politano_Pinton_ind} 
using LES, our result appears to be the first time that it has 
been obtained in a direct numerical simulation of the low-$\Rm$ MHD 
turbulence. Although no element of surprise was present 
here, we consider it reassuring to have dotted this particular i. 

\subsection{Turbulent Induction at High $\Rm$} 
\label{sec_highRm}

As $\Rm$ increases, 
the quasistatic approximation \eqref{quasist} ceases to be valid, 
the nonlinear terms in \eqref{mag_ind} are no longer negligible and 
the dependence $f(\Rm)$ becomes nontrivial. 
\Figref{fig_E}(b) shows how at $\Rm>2$ it flattens until, 
around $\Rm\sim200$, the fluctuation dynamo sets in, 
overwhelms the turbulent induction and brings the magnetic 
energy into a saturated state determined not by 
$\Rm$ but by the nonlinear back reaction 
($\dBsq\sim\usq$).\footnote{Note that while the scaling of $\dBsq$ with 
$\Rm-\Rmc$ just above the transition to dynamo is an interesting 
theoretical question \cite{Petrelis_Fauve_crit}, 
determining this scaling numerically is not currently 
possible because of the extreme long-time fluctuations 
close to criticality and the consequent need for unaffordably 
long runs.} The close agreement between the magnetic 
energies for the Laplacian and hyperviscous runs with different 
$\Re$ suggests that these results are converged in $\Re$. 

While one might dwell on the question of what the asymptotic form of $f(\Rm)$ 
for large $\Rm$ should be 
\citeaffixed{Moffatt,Parker63_1,Saffman,BCLow,Vainshtein_Cattaneo,Cattaneo_etal}{e.g.,}, 
it is perhaps reasonable to ask first whether, in view 
of our claim that the fluctuation dynamo is unavoidable at 
sufficiently large $\Rm$, the problem is meaningfully posed. 
The short answer is, obviously, no. However, there is a useful way, 
already intimated at the beginning of \secref{sec_ind}, in 
which the turbulent induction problem can be posed at high $\Rm$. 

Firstly, if $\vB_0$ is interpreted as the 
dynamo-generated field at the resistive scale and above, 
we may inquire into the behaviour of the magnetic fluctuations 
below the resistive scale, $l\ll\lres$. Since 
the inertial-range motions at these scales 
have a shorter correlation time than at $l\sim\lres$, 
we can, indeed, treat this as a problem with a constant mean field 
and use \eqref{mag_ind}. 
The quasistatic theory is still valid because the diffusion 
term is dominant. Thus, the \citeasnoun{Golitsyn} spectrum 
is now recovered as the subresistive tail of the 
magnetic-energy spectrum \cite{Moffatt}. In numerical simulations, 
this is hard to check at current resolutions, but we can compare the 
magnetic spectra in our simulations with the quasistatic prediction 
\eqref{sp_quasist} using the full numerically obtained form of $E(k)$ 
and replacing $\Bo\to\Bsq^{1/2}$. There is, indeed, a fit below some 
sufficiently small (resistive) scale, which predictably decreases 
with increasing $\Rm$. 

Secondly, if the dynamo amplifies the magnetic field at the outer 
scale or above (this is now our mean field $\vB_0$),
one might ask how much magnetic energy this will 
generate via the turbulent induction 
in the part of the inertial range that lies above the resistive 
scale, $\lf\gg l\gg\lres$. 
In \eqref{mag_ind}, the diffusive term can now be ignored. 
It seems then to be a plausible argument that, if the nonlocal energy 
transfer from the outer-scale field is important at all, 
its effect on the inertial-range magnetic fluctuations 
can be found by balancing the ``source'' term containing $\vB_0$ 
with the nonlinear terms, which represent the local interactions between $\vu$ 
and $\dvB$. This gives \cite{Ruzmaikin_Shukurov} 
\bea
\dB_l\sim\Bo \quad\Rightarrow\quad 
M(k)\sim \Bo^2\, k^{-1}.
\label{sp_RS}
\eea
The same spectrum and the 
consequent scaling $\dBsq\sim(\ln\Rm)\Bo^2$ were obtained in 
several closure calculations assuming a weak mean field and 
no dynamo \cite{Kleeorin_Rogachevskii_Ruzmaikin,KR_minus1,Kleeorin_Mond_Rogachevskii}
(see, however, an argument by \citeasnoun{Moffatt}, 
based on the mathematical analogy between the magnetic field 
and vorticity and leading to a $k^{1/3}$ spectrum). Note that all 
this only applies in the limit $\dBsq\ll\usq$ (which 
is where our simulations are; see \figref{fig_E}(b)). Otherwise, 
dynamical effects, such as the Alfv\'enization of the turbulence, 
will be important and determine the shape of the saturated state. 

The scaling \eqref{sp_RS} can only be realized if an outer-scale magnetic field 
really exists and if the tangling of this field by turbulence is not 
superceded by an inertial-range dynamo, as discussed in 
\secref{sec_faqs}. Thus, it is only a possibility and, indeed, 
a signature property, if the fluctuation dynamo found by us is, 
in fact, an outer-scale dynamo. As we explained in \secref{sec_spectra}, 
our numerical simulations are not sufficiently asymptotic 
to determine the spectrum of magnetic fluctuations. 
The apparent tendency towards a negative spectral slope 
reported in \secref{sec_spectra} may be a telling sign in view 
of the theoretical result \eqref{sp_RS}. We have plotted  
the reference $k^{-1}$ slopes in \figref{fig_spRm} and 
\figref{fig_spPm}(e,f). We leave it to the reader's 
judgement to decide if this scaling might, indeed, 
be emerging there.  

An important observation must be made in this context. 
The saturated magnetic-energy spectra in the subcritical 
runs with an imposed 
weak mean field turn out, after normalization, to be 
{\em exactly} the same as the normalized spectra of the 
corresponding decaying runs without the mean field 
(see \figref{fig_HM}(a)). Furthermore, as reported in 
\secref{sec_spectra}, no qualitative change occurs in the 
magnetic-energy spectrum as the dynamo threshold is crossed. 
This seems to tell us that the same mechanism is responsible 
for setting the shape of the spectrum of the magnetic fluctuations 
induced by a mean field and of the decaying or growing such 
fluctuations in the absence of a mean field. 

\section{Conclusions}
\label{sec_conc}

Let us reiterate the main numerical results and theoretical 
points presented above. 

\begin{itemize}

\item Fluctuation dynamo exists in nonhelical randomly forced 
homogeneous turbulence of a conducting fluid with low magnetic Prandtl 
number (\secref{sec_Rmc}). 
The critical magnetic Reynolds number for this dynamo 
is at most three times larger than for $\Pm\ge1$: 
defined by \eqref{Re_def}, it is $\Rmc\lesssim 200$ 
for $\Re\gtrsim6000$, although there is a larger 
peak value at a somewhat smaller $\Re$. 

\item The nature of the dynamo and its stability curve 
$\Rmc(\Re)$ are different from the dynamo obtained 
in simulations and liquid-metal experiments with a
mean flow (\secref{sec_meanflow}). 

\item The physical mechanism that enables the sustained 
growth of magnetic fluctuations in the low-$\Pm$ regime 
is unknown. It is not as yet possible to determine numerically 
whether the fluctuation dynamo is driven by the inertial-range motions 
at the resistive scale 
and consequently has a growth rate $\propto\Rm^{1/2}$ in the limit 
$\Rm\to\infty$, or rather a constant growth rate comparable to 
the turnover rate of the outer-scale motions 
(\secref{sec_faqs}). 

\item The magnetic-energy spectra in the low-$\Pm$ 
regime are qualitatively different from the $\Pm\ge1$ case 
and appear to develop a negative spectral slope, which 
may be consistent with $k^{-1}$, but cannot be definitively 
resolved (\secref{sec_spectra}). 
The spectra of the growing field are similar to those 
for the decaying field at lower $\Rm$ and to the 
saturated spectra of the induced magnetic energy in the 
presence of a weak mean field (\secref{sec_highRm}). 

\item At very low $\Rm$, the magnetic fluctuations induced 
via the tangling by turbulence of a weak mean field 
are well described by the quasistatic approximation. 
The $k^{-11/3}$ spectrum is confirmed (\secref{sec_lowRm}). 

\end{itemize}

While these results leave a frustrating number of questions 
unanswered and do not entirely clear up the confusion over 
the exact nature of the low-$\Pm$ dynamo, they do 
at least confirm that the object of this confusion exists. 
It is important to check in independent numerical experiments 
both that our conclusions hold and whether the value
of the dynamo threshold obtained by us is universal. 
Some promising numerical experiments aimed at elucidating 
the nature of the dynamo and the role of the mean flow 
are proposed in \secref{sec_meanflow} and \secref{sec_faqs}. 

The main conclusion of this work is the confirmation that 
nature will always find a way to make magnetic field where 
turbulence of a conducting fluid is present. In the astrophysical 
context, the low-$\Pm$ fluctuation dynamo is particularly 
(although by no means solely) important in the context 
of solar magnetism. While there is ample observational evidence that
small-scale magnetic fluctuations pervade the solar photosphere 
\citeaffixed{Dominguez_etal,Socas_Lites,Trujillo_etal,Solanki_etal_review}{e.g.,}, 
the numerical evidence used to argue that the turbulent fluctuation 
dynamo is responsible has thus far been based on 
numerical simulations with $\Pm\ge1$ 
\cite{Cattaneo,Cattaneo_Emonet_Weiss,Bushby}. 
The existence of the low-$\Pm$ fluctuation dynamo that we 
have established for an idealized homogeneous MHD turbulence 
gives us confidence that attempts to demonstrate self-consistent 
magnetic-field amplification in more realistic 
simulations of solar convection with $\Pm\ll1$ will eventually 
prove successful. 

\ack

We thank A.~Brandenburg, B.~Dubrulle, S.~Fauve, C.~Forest, N.~Haugen, D.~Hughes, N.~Kleeorin, 
P.~Mininni, J.~Papaloizou, J.-F.~Pinton, F.~Plunian, Y.~Ponty, F.~Rincon, I. Rogachevskii, 
A.~Shukurov and N.~Weiss for valuable discussions at various stages of this project.
We are grateful to V.~Decyk (UCLA) who has kindly provided his FFT libraries
from the UPIC framework. 
A.A.S.\ was supported by a PPARC Advanced Fellowship. 
He also thanks the USDOE Center for Multiscale Plasma Dynamic for travel 
support and the UCLA Plasma Group for its hospitality. 
A.B.I.\ was supported by the USDOE Center for Multiscale Plasma Dynamics. 
T.A.Y.\ was supported by a UKAFF Fellowship and the Newton Trust. 
Simulations were done on UKAFF (Leicester), NCSA (Illinois) and 
the Dawson cluster (UCLA Plasma Group). 

\section*{References}

\bibliographystyle{jphysicsB}
\bibliography{sicmpy_NJP}

\end{document}